\documentclass[%
 reprint,
nofootinbib,
 amsmath,amssymb,
 aps,
]{revtex4-2}

\usepackage{graphicx}
\usepackage{dcolumn}
\usepackage{bm}
\usepackage{siunitx}
\usepackage[ruled,vlined]{algorithm2e}

\begin{document}

\preprint{APS/123-QED}

\title{Energy-dependent implementation of secondary electron emission models in continuum kinetic sheath simulations}

\author{Kolter Bradshaw}
 \email{kolterb@vt.edu}
 \affiliation{%
    Kevin T. Crofton Department of Aerospace and Ocean Engineering, Virgina Tech
}%
\author{Bhuvana Srinivasan}%
 \email{srinbhu@uw.edu}
\affiliation{%
 University of Washington, Seattle WA 98195
}%
\affiliation{%
 Kevin T. Crofton Department of Aerospace and Ocean Engineering, Virgina Tech
}%

\date{\today}

\begin{abstract}
    The plasma-material interactions present in multiple fusion and propulsion concepts between the flow of plasma through a channel and a material wall drive the emission of secondary electrons. This emission is capable of altering the fundamental structure of the sheath region, significantly changing the expected particle fluxes to the wall. The emission spectrum is separated into two major energy regimes, a peak of elastically backscattered primary electrons at the incoming energy, and cold secondary electrons inelastically emitted directly from the material. The ability of continuum kinetic simulations to accurately represent the secondary electron emission is limited by relevant models being formulated in terms of monoenergetic particle interactions which cannot be applied directly to the discrete distribution function. As a result, rigorous implementation of energy-dependent physics is often neglected in favor of simplified, constant models. We present here a novel implementation of semi-empirical models in the boundary of continuum kinetic simulations which allows the full range of this emission to be accurately captured in physically-relevant regimes. 
\end{abstract}

\maketitle


\section{Introduction}

In a wide variety of applications, such as Hall thrusters \cite{goebel2023} and a variety of fusion devices including the tokamak \cite{artsimovich1972}, plasmas are constrained to flow within a material channel. Impact of particles on the walls leads to material degradation, damaging the device and causing contamination of the plasma \cite{stangeby2000}. Transfer of plasma particles to the wall is regulated by the plasma sheath, a region of decreasing potential which balances the electron and ion fluxes to the surface \cite{robertson2013}. Simulations of the plasma sheath are a valuable tool for making accurate predictions of factors important to the design of applications, such as particle fluxes to the wall and the plasma regime in the boundary layer \cite{hua1994,taccogna2005,cagas2017_sheath,li2023,skolar2023}. The presence of emission mechanisms in the sheath has the potential to change behavior significantly from classical models \cite{hobbs1967}, causing transition from the monotonic `classical' sheath mode to the non-monotonic space-charge limited (SCL) or inverse modes.\par
Continuum kinetic models of plasmas forego direct particle representation in favor of evolving the discretized distribution function. This allow simulations to capture small-scale kinetic physics which is often lost by fluid models, while avoiding the statistical noise that poses problems for particle-in-cell (PIC) methods \cite{birdsall1991}. When applied to the situation of the classical plasma sheath which forms at the boundary of a plasma and a material surface, continuum kinetic models allow for accurate simulation of the sheath region and instabilities \cite{cagas2017_sheath, cagas2018_diss}. However, the addition of secondary electron emission (SEE) physics poses several unique challenges to these models. The ability to simulate SEE rigorously in continuum kinetic codes is heavily restricted by the need to apply emission models, which describe monoenergetic particle interactions, to the discrete distribution function.\par
Past treatments of the emitting plasma sheath have typically modeled the emission using simplified approaches. Most theory assumes a thermionic-type emission, treating the emitted population as a uniform cold Maxwellian distribution \cite{hobbs1967,hall1976,stangeby1984,schwager1993}. A similar treatment of the emitted population in continuum kinetic codes has been used for  demonstrating the different sheath modes which occur as the magnitude of emission increases \cite{campanell2016}. While useful for demonstrating different emission regimes and for applications where thermionic emission is present, such a treatment is insufficient to capture the complex features of the secondary electron emission driven by particle impact on a material. Several different emission mechanisms contribute to total emission \cite{jensen2018}, each producing a unique distribution of emitted particles. For simulations to be useful as predictive tools for physical applications, a more rigorous implementation is required.\par
More complete descriptions of the emission have been attempted in PIC codes. These codes are able to sample the emitted distribution for individual impact events using Monte Carlo methods \cite{furman2003,taccogna2004,azzolini2019}. Continuum kinetic codes are not capable of this approach as they do not handle individual particle interactions with the wall, and therefore struggle to represent the full range of emission behaviors across all incoming energies. These restrictions often lead to oversimplified or constant implementations of complex physics, which fail to capture the dynamic nature of actual physical applications. The purpose of the implementation shown here is to bring the full range of energy-dependent physics into a flexible continuum kinetic implementation. Presented in this work is an overview of different approaches to modeling the SEE, discussion of their limitations, and novel implementation of phenomenological models in a continuum kinetic framework. The implementation is generally extendable to any arbitrary material and plasma regime.\par

\section{Emitting Sheaths}

The plasma sheath forms due to the greater mobility of the electrons over ions. Greater electron current at the surface causes a negative potential to develop, which drives the formation of a positive space-charge region near the material. Ions are accelerated while electrons are repelled, leading to an equalization of the fluxes at the wall. The characteristic length scale in the sheath is the Debye length,
\begin{equation}
    \lambda_D = \sqrt{\frac{\varepsilon_0T_e}{n_eq_e^2}},
\end{equation}
where $\varepsilon_0$ is the vacuum permittivity, $T_e$ is the electron temperature, $n_e$ is the electron number density, and $q_e$ is electron charge. The length of the sheath region is typically on the order of some tens of Debye lengths. At the sheath entrance, ions must be accelerated to the Bohm speed $u_B$. Several formulations of this resulting Bohm sheath criterion are possible \cite{bohm1949,riemann1991,li2022}, but the one used for the purposes of this work has the form \cite{tang2017}
\begin{equation}
    u_i\geq u_B = \sqrt{\frac{\gamma_eT_e + \gamma_iT_i}{m_i}},
    \label{eq:bohm}
\end{equation}
where $u_i$ is the ion drift speed in the sheath, and $\gamma$ is the heat capacity ratio.\par
The above overview of the theory assumes the wall perfectly absorbs any particles which impact it. In physical cases, however, impact of energetic ``primary" electrons leads to the emission of the so-called ``secondary" electrons from the wall material into the sheath, considerably impacting the resulting physics. The emission of secondary electrons also occurs with the impact of primary ions. We define the secondary electron yield (SEY) to be the number of secondary particles emitted per primary particle impact. The SEY of a surface depends on material properties and the energy and angle of the primary particle. For low SEY below unity, the sheath will remain ``classical"; that is, monotonic with a negative wall potential. At some critical SEY near unity, the sheath potential becomes non-monotonic and transitions to a space-charge limited (SCL) sheath \cite{hobbs1967,schwager1993}. Theory further predicts that when the ratio of emitted flux to flux from the bulk plasma exceeds unity, ion collisional effects will drive the SCL sheath to a reverse sheath where the wall potential is positive relative to the sheath entrance \cite{campanell2013,campanell2016}.\par
Modifications can be made to the sheath theory to account for the presence of particle emission \cite{hobbs1967,schwager1993,campanell2013,sheehan2014}. These models, however, generally invoke simplifying assumptions about the emitted particle distribution, considering it to be constant, cold, and half-Maxwellian, typical of thermionic probe emission. The emission of a material under particle impact, however, is significantly more dynamic and varied. Measurements of the SEE spectrum for various materials under electron impact show three distinct populations of emitted particles \cite{jensen2018}. A peak is located at the incident energy representing particles which reflect approximately elastically off the wall. A peak at low energy represents the so-called ``true" secondary electrons emitted from the material. The low emission region between these two contains the inelastically reflected, or rediffused particles, which penetrate the surface but undergo several scattering events and are reemitted at lower energy.

\section{Numerical Model}

\subsection{The Kinetic Equations}

The core kinetic model is the Vlasov-Maxwell-Fokker-Planck (VM-FP) system of equations, coupling Maxwell's equations with the Boltzmann equation
\begin{equation}
    \frac{\partial f_s}{\partial t} = -\mathbf{v}\cdot \frac{\partial f_s}{\partial\mathbf{x}} - \frac{q_s}{m_s}(\mathbf{E} - \mathbf{v}\times\mathbf{B})\cdot\frac{\partial f_s}{\partial\mathbf{v}} + \bigg(\frac{\partial f_s}{\partial t}\bigg)_c,\label{eq:vlasov}
\end{equation}
where $f_s$ is the particle distribution function for species $s$. In this work, Coulomb collisions are represented by the Lenard-Bernstein (LBO) collision operator \cite{dougherty1964}. This operator takes the form
\begin{equation}
    \bigg(\frac{\partial f_s}{\partial t}\bigg)_c = \sum_r\nu_{sr}\frac{\partial}{\partial\mathbf{v}}\cdot\left[(\mathbf{v}-\mathbf{u}_{sr})f_s + v_{t,sr}^2\frac{\partial f_s}{\partial\mathbf{v}}\right],
\end{equation}
where $\nu_{sr}$ is the collision frequency with species $r$ and $v_{t,sr}$ is the cross-flow thermal velocity.\par

In this work, the discretization of the distribution function and evolution of Eq.~\ref{eq:vlasov} is done using the discontinuous Galerkin (DG) numerical method \cite{cockburn2001} with the \texttt{Gkeyll} software \cite{gkyldocs}. It is nevertheless our intent that this algorithm be generally applicable to variety of computational methods. Therefore, while some of the DG details will be given where necessary, the work shown here should be valid for any numerical discretization technique. The \texttt{Gkeyll} code employs a multistage Runge-Kutta (RK) scheme of successive forward Euler steps, described in Alg.~\ref{algo:rk_step}, but the general procedure of calculating the right-hand side (RHS) terms of Eq. \ref{eq:vlasov} and advancing them by some time step is common to all relevant schemes.\par

\begin{algorithm}
    \SetAlgoLined
    \SetKwFunction{RK}{rkStage}
    \SetKwFunction{bc}{bcCalc}
    \tcp{Calculation of VM-FP RHS terms}
    $\mathcal{L} = \frac{\partial f^n}{\partial t}$\\
    \tcp{Forward Euler update}
    $\mathcal{F} = f^n + \Delta t\mathcal{L}$\\
    \tcp{RK stage update}
    $f^{n+1} = $ \RK{$f^n, \mathcal{F}$}\\
    \tcp{Calculation of ghost cell distribution from skin cell distribution}
    $f^{g,n+1}=$ \bc{$f^{s,n+1}$}
    \caption{Continuum kinetic update algorithm}
    \label{algo:rk_step}
\end{algorithm}

Of particular relevance to this work are the skin and ghost cells. The ghost cells are a set of cells place outside the edge of the domain which are set by the boundary condition in the previous time step, and are used in the RHS update of the interior cell layer bordering the domain edge (the ``skin" cells) during the DG update. In most cases dealing with particle emission, the ghost cell distribution calculated by the boundary condition will depend on the incoming distribution in the skin cells  \cite{cagas2020}.\par
A thorough outline of the full DG discretization of the VM-FP system is presented in  \cite{hakim2014,juno2018,hakim2020,juno2020,HakimJuno:2020}. As this paper deals extensively with transformations of the discrete distribution function, the weak formulation is described in the following section.\par

\subsection{The Discrete Distribution Function}

The distribution function $f(\mathbf{x}, \mathbf{v})$ and the discrete distribution function $f_h(\mathbf{x}, \mathbf{v})$ are related through \emph{weak equality} $f(\mathbf{x}, \mathbf{v})\circeq f_h(\mathbf{x}, \mathbf{v})$, that is
\begin{equation}
    \begin{split}
        \int_{\mathcal{X}}\int_{\mathcal{V}} f(\mathbf{x}, \mathbf{v})\psi_t(\mathbf{x}, \mathbf{v})d\mathbf{x}d\mathbf{v}\\
        = \int_{\mathcal{X}}\int_{\mathcal{V}} f_h(\mathbf{x}, \mathbf{v})\psi_t(\mathbf{x}, \mathbf{v})d\mathbf{x}d\mathbf{v},
    \end{split}
    \label{eq:weak_equality}
\end{equation}

where $\psi_t(\mathbf{x}, \mathbf{v})$ is the test function. To evaluate the discrete distribution function, we do a coordinate transformation from phase space ($\mathcal{X}, \mathcal{V}$) with ($\mathbf{x}, \mathbf{v}$) coordinates to logical space $\mathcal{I}$ with ($\boldsymbol{\eta}_{\mathbf{x}}, \boldsymbol{\eta}_{\mathbf{v}}$) coordinates spanning the range $-1$ to $1$ in each dimension in each cell,
\begin{equation}
    x_i(\eta_i) = \eta_i\frac{\Delta x_i}{2} + x_{ci},\quad -1\leq\eta_i\leq 1.
    \label{eq:logical_space}
\end{equation}
Here, $x_{ci}$ is the cell center value, and $\Delta x_i$ is the cell width. We take the discrete distribution function in each cell to be the sum of $N_b$ expansion coefficients ($\hat{f}=[\hat{f}_0, \hat{f}_1, ..., \hat{f}_{N_b-1}]$) and basis functions ($\hat{\psi}(\boldsymbol{\eta}_{\mathbf{x}}, \boldsymbol{\eta}_{\mathbf{v}})=[\hat{\psi}_0(\boldsymbol{\eta}_{\mathbf{x}}, \boldsymbol{\eta}_{\mathbf{v}}), \hat{\psi}_1(\boldsymbol{\eta}_{\mathbf{x}}, \boldsymbol{\eta}_{\mathbf{v}}), ..., \hat{\psi}_{N_b-1}(\boldsymbol{\eta}_{\mathbf{x}}, \boldsymbol{\eta}_{\mathbf{v}})]$).

Thus, the representation of the distribution function in cell $i$ in configuration space and $j$ in velocity space is

\begin{equation}
    \begin{aligned}
        f^{ij}(\mathbf{x}, \mathbf{v}) &\circeq f^{ij}_h(\mathbf{x}, \mathbf{v}),\quad\mathbf{x},\mathbf{v}\in\mathcal{X}^i,\mathcal{V}^j\\
        & = \sum_{k=0}^{N_b-1}\hat{f}^{ij}_{k}\hat{\psi}_k(\boldsymbol{\eta}_{\mathbf{x}}, \boldsymbol{\eta}_{\mathbf{v}}).
    \end{aligned}
\end{equation}

This results in Eq.~\ref{eq:weak_equality} becoming
\begin{equation}
    \mathbf{M}\hat{f}^{ij} = \int_{\mathcal{I}} f\big(\mathbf{x}(\boldsymbol{\eta}_{\mathbf{x}}), \mathbf{v}(\boldsymbol{\eta}_{\mathbf{v}})\big)\psi_t(\boldsymbol{\eta}_{\mathbf{x}}, \boldsymbol{\eta}_{\mathbf{v}})d\boldsymbol{\eta}_{\mathbf{x}}d\boldsymbol{\eta}_{\mathbf{v}},
    \label{eq:logical_weak}
\end{equation}
where 
$$\mathbf{M} = \int_{\mathcal{I}}\hat{\psi}(\boldsymbol{\eta}_{\mathbf{x}}, \boldsymbol{\eta}_{\mathbf{v}})\psi_t(\boldsymbol{\eta}_{\mathbf{x}}, \boldsymbol{\eta}_{\mathbf{v}})d\boldsymbol{\eta}_{\mathbf{x}}d\boldsymbol{\eta}_{\mathbf{v}}$$
is the mass matrix, and the expressions $\mathbf{x}(\boldsymbol{\eta}_{\mathbf{x}})$, $\mathbf{v}(\boldsymbol{\eta}_{\mathbf{v}})$ are taken from Eq.~\ref{eq:logical_space}. In \texttt{Gkeyll}, we choose the same orthonormal polynomial basis set for both the basis functions $\hat{\psi}(\boldsymbol{\eta}_{\mathbf{x}}, \boldsymbol{\eta}_{\mathbf{v}})$ and the test functions $\psi_t(\boldsymbol{\eta}_{\mathbf{x}}, \boldsymbol{\eta}_{\mathbf{v}})$. This orthonormality causes the mass matrix $\mathbf{M} = \mathbf{I}_b,$ where $\mathbf{I}_b$ is the $N_b\times N_b$ identity matrix. The integral on the right-hand side of of Eq.~\ref{eq:logical_weak} can be evaluated numerically to obtain the DG expansion coefficients of the discrete distribution function.

\section{Secondary Electron Emission Models}

The models discussed in this work deal exclusively with electron-impact SEE. Ion-impact SEE is also possible in regimes of high energy ions, and models for these interactions will be addressed in detail in future related work. The fundamental algorithms here, however, are agnostic to whether the impacting and emitted species are the same. We take the subscript $\alpha$ to denote the particle species impacting the wall, and $\beta$ to denote the species being emitted. We further denote the energy of an impacting particle $E'$ with angle cosine $\mu'$, and the emitted particle energy $E$ with angle cosine $\mu$. Energetic quantities are all in electronvolts (eV). These parameters are related directly to the particle velocity.
\begin{equation}
    E_s = \frac{1}{2}\frac{m_s}{q_0}\mathbf{v}\cdot\mathbf{v},
\end{equation}
\begin{equation}
    \mu = \frac{\mathbf{n}\cdot\mathbf{v}}{|\mathbf{v}|},
\end{equation}
with $q_0$ being elementary charge, and $\mathbf{n}$ being the unit vector normal to the wall. Typically in this work, emission models will be formulated in terms of $E$ and $\mu$, while distribution functions will be expressed in terms of $\mathbf{v}$.\par
We further define the secondary electron yield (SEY) to be the ratio of the emitted particle flux to the impacting particle flux $\delta = \Gamma_{\beta}^-/\Gamma_{\alpha}^+$. For this work, the positive direction is defined as into the wall, the negative direction as out of the wall.\par
In this work we treat the populations separately, denoting the elastic yield as $\delta_e$, and the true secondary yield as $\delta_{ts}$. The rediffusion, $\delta_r$, will not be implemented by this work as it minimally contributes in most situations to the overall emitted population and presents unique modeling difficulties. Typically, at lower energies backscattering dominates emission, while true secondary emission dominates high energy situations. As will be shown, neither stand completely independent in any regime.

\begin{figure}
    \includegraphics[width=1.0\linewidth]{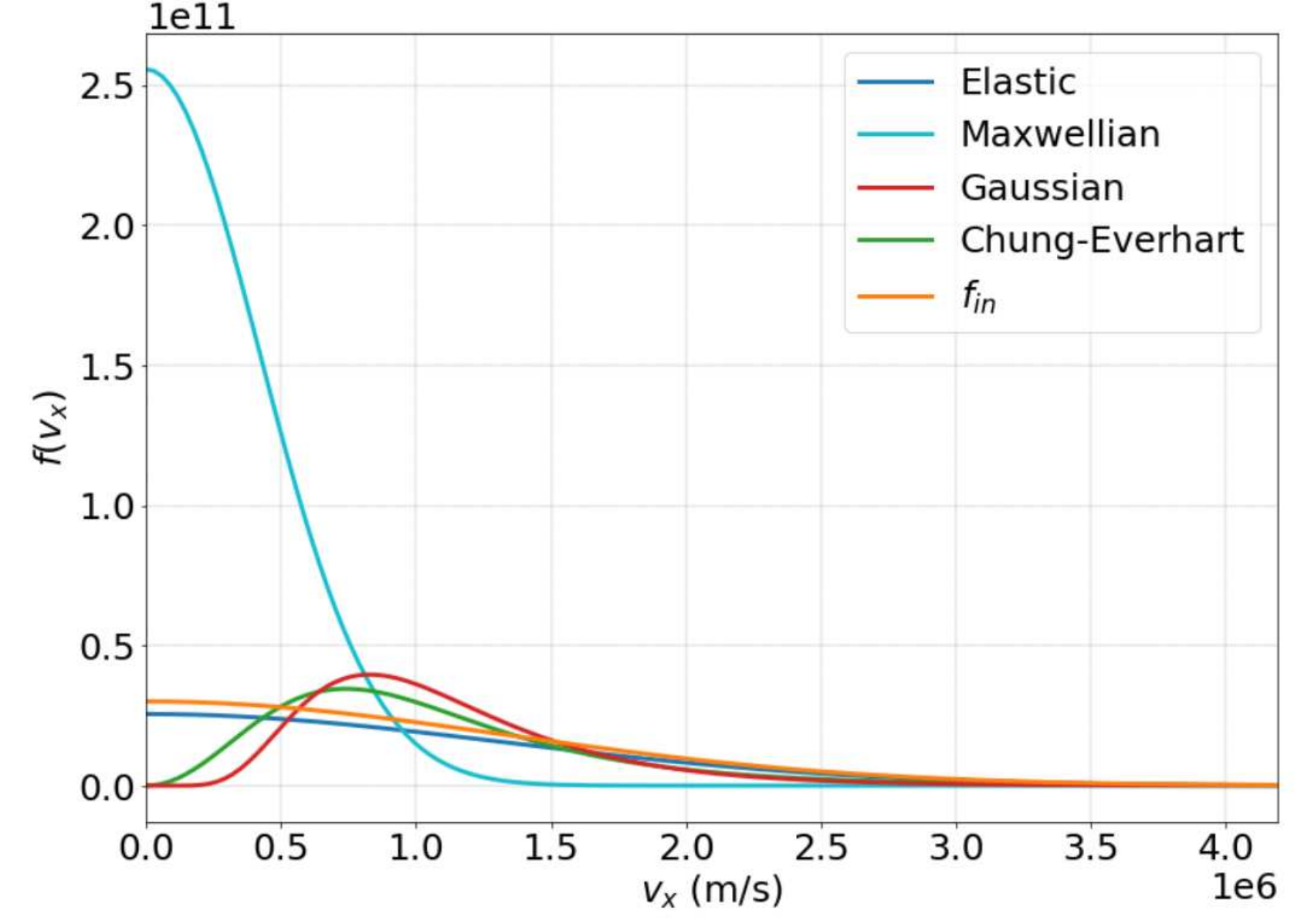}
    \caption{\label{fig:constant_spectra} Comparison of different secondary electron emission spectra when $\delta=0.85$. While the flux ratio is constant across each case, the resulting densities are quite different. When the emitted particles are at low energy, they pile up significantly more at the wall than emission spectra with high velocity tails.}
\end{figure}

\subsection{Constant Emission}

The simplest means of extending constant emission to an inelastic emission spectrum is to scale a representative curve by a normalization factor $C$, such that

\begin{equation}
    f^g_{\beta}(\mathbf{x}, \mathbf{v}) = Cf_{ts}(\mathbf{v}).
\end{equation}

Here $f_{ts}(\mathbf{v})$ is the function for the emission spectrum, and $C$ is a normalization factor that ensures the flux ratio is correct. The simplest approach is elastic emission, setting $C=\delta_{e}$ and $f_{ts}(\mathbf{v}^-) = f_{\alpha}^s(\mathbf{x}_{wall}, \mathbf{v}^+)$.\par
An inelastic implementation for the true secondary population is more sophisticated, using the impacting flux to scale some emission spectrum function. From the definition of $\delta_{ts}$ using the ratio of flux/first moment,

\begin{equation}
    C = \frac{\delta_{ts}\Gamma_{\alpha}^+}{\int_{\mathcal{V}^-}\mathbf{n}\cdot\mathbf{v}f_{ts}(\mathbf{v})d\mathbf{v}}.
    \label{eq:norm_calc}
\end{equation}

The denominator can be solved analytically for a chosen $f_{ts}(\mathbf{v})$ spectrum. For a Maxwellian
\begin{equation}
    f_{ts}(\mathbf{v}) = \exp{\bigg(-\frac{\mathbf{v}\cdot\mathbf{v}}{2v_{emit}^2}\bigg)},
\end{equation}
this expression yields

\begin{equation}
    C_M = \frac{\delta_{ts}\Gamma_{\alpha}^+}{(2\pi)^{\frac{d_v-1}{2}}v_{emit}^{d_v+1}},
\end{equation}
for dimensions $d_v$ in velocity space.

Here, $v_{emit} = \sqrt{\frac{k_BT_{emit}}{m_e}}$ will typically be chosen with a cold $T_{emit}$ to match standard true secondary electron emission behavior.

Other spectrum function choices are available and supported by emission data, such as the Gaussian function \cite{scholtz1996}
\begin{equation}
    f_{ts}(\mathbf{v}) = \exp{\bigg(-\frac{\ln{\big(\frac{E(\mathbf{v})}{E_0}\big)}^2}{2\tau^2}\bigg)},
\end{equation}
where $E_0$ and $\tau$ are fitting parameters, and the Chung-Everhart \cite{chung1974} model
\begin{equation}
    f_{ts}(\mathbf{v}) = \frac{E(\mathbf{v})}{(E(\mathbf{v}) + \phi)^4},
\end{equation}
where $\phi$ is the material work function (or electron affinity for insulators). The fluxes of these functions do not give analytical solutions which scale cleanly into higher dimensions (see Section VI-A), but in 1V the solutions are

\begin{equation}
    C_G = \frac{\delta_{ts}\Gamma_{\alpha}^+m_{\beta}}{\sqrt{2\pi}q_0E_0\tau\exp{(\tau^2/2)}},
\end{equation}

and 
\begin{equation}
    C_{CE} = \frac{6m_{\beta}}{q_0}\phi^2\delta_{ts}\Gamma_{\alpha}^+,
\end{equation}
for the Gaussian and Chung-Everhart functions, respectively.\par
Comparisons of how the choice of spectrum changes the emission for the same yield value are shown in Fig.~\ref{fig:constant_spectra}. Despite having the same flux ratio, the resulting distribution is vastly different. Particles at high velocity contribute more to emission than particles at low velocity, so the Maxwellian centered at zero emits at far greater density than the other curves.\par
There are two drawbacks in using either constant elastic or inelastic emission, or some combination of the two. First, as these $\delta$ values are constant across space, low energy particles nonphysically contribute to the cold secondaries, and conversely there is nonphysical backscattering of high energy particles back into the sheath. Second, any constant application of the yield across space also results in a necessarily constant yield in time; and as such, the implementation does not capture any feedback between the emission and the sheath.

\begin{figure}
    \includegraphics[width=1.0\linewidth]{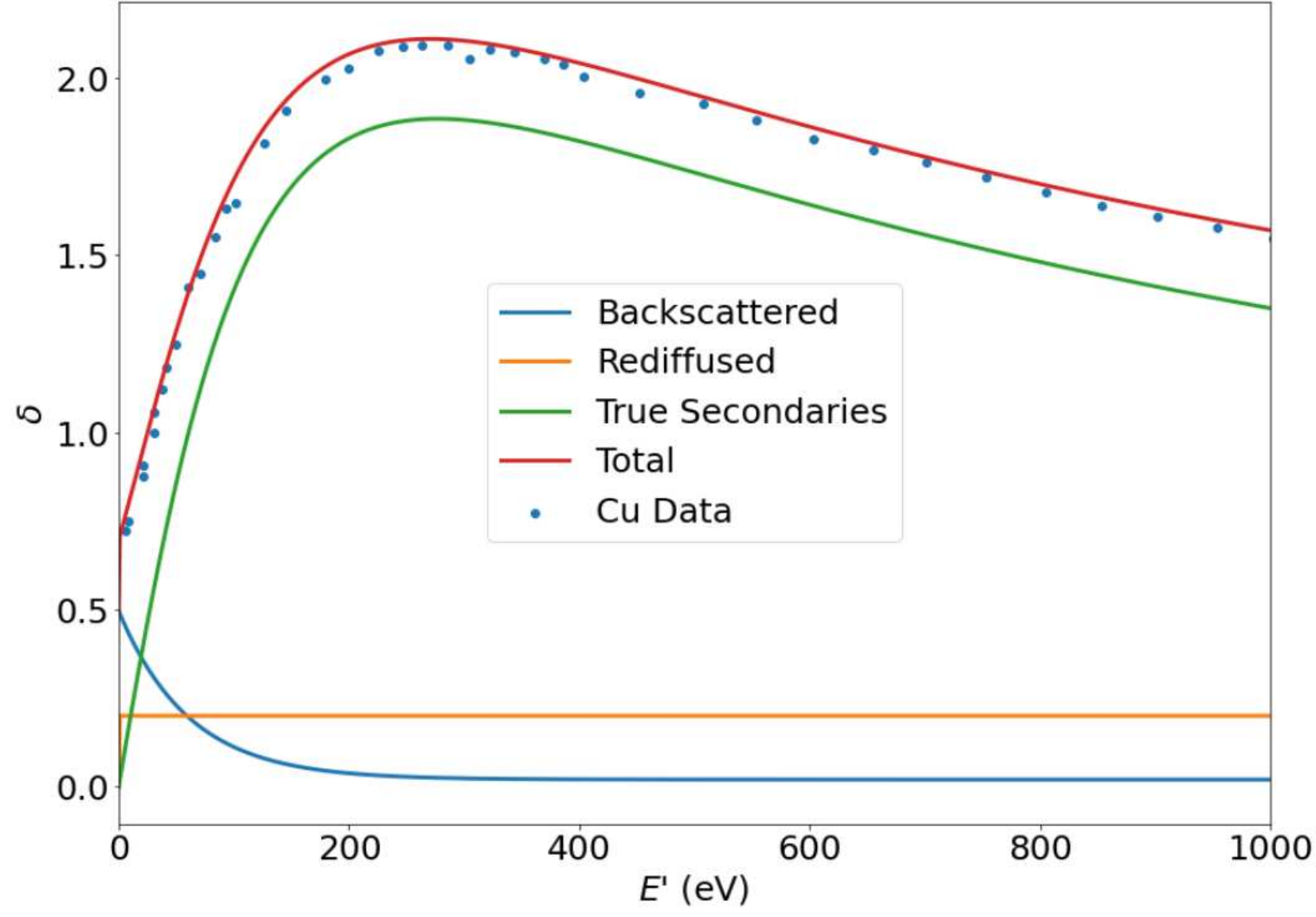}
    \includegraphics[width=1.0\linewidth]{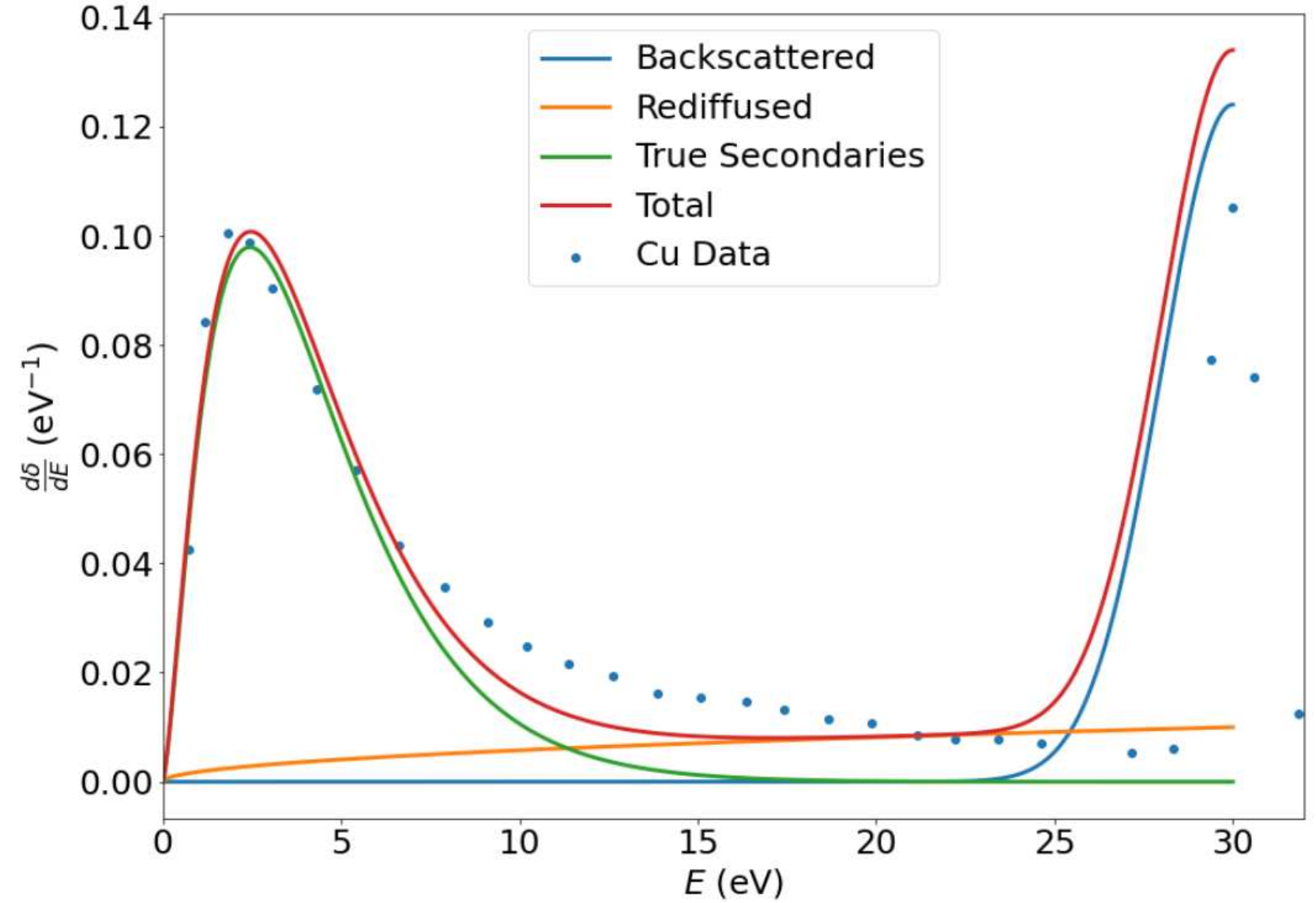}
    \caption{\label{fig:fp_fit} Emission data \cite{baglin2001} for copper yield (top) and $\SI{30}{eV}$ beam emission spectrum (bottom) with Furman-Pivi fits \cite{furman2002} to the three populations of secondary electrons. There is a near-elastic peak at the beam energy which corresponds to the backscattering, while regardless of beam energy the location of the true secondary peak will remain approximately the same.}
\end{figure}

\subsection{Furman-Pivi Model}

The Furman-Pivi model \cite{furman2002} gives the SEE emission spectrum for a monoenergetic beam of impacting electrons; thus, $\alpha$ and $\beta$ are both electrons. There are two parts to the Furman-Pivi model. The SEY which gives the number of particles emitted per incident particle,
\begin{equation}
    \delta_{tot}(E', \mu') = \delta_e(E', \mu') + \delta_r(E', \mu') + \delta_{ts}(E', \mu'),
\end{equation}

and the spectrum which describes how the emitted particles are distributed across outgoing energy,

\begin{equation}
    \begin{split}
        \frac{\partial\delta_{tot}}{\partial E}\bigg(E,E',\mu'\bigg) = \frac{\partial\delta_e}{\partial E}\bigg(E,E',\mu'\bigg)\\
        + \frac{\partial\delta_{r}}{\partial E}\bigg(E,E',\mu'\bigg) + \frac{\partial\delta_{ts}}{\partial E}\bigg(E,E',\mu'\bigg).
    \end{split}
\end{equation}

These fits are shown for copper data in Fig.~\ref{fig:fp_fit}. As mentioned, we are neglecting the rediffused population represented by $\delta_r$, $\frac{\partial\delta_{r}}{\partial E}$. Additionally, while the Furman-Pivi model give a spectrum for the backscattered population, we note that the backscattering can be approximated well as purely elastic, and thus drop $\frac{\partial\delta_e}{\partial E}$ as a full spectrum implementation. Discussion here will focus therefore on the implementation of the true secondary population. The separate handling of elastic backscattering terms is discussed in Section IV-C.\par
The Furman-Pivi fit to the SEY curve is
\begin{equation}
    \delta_{ts}(E', \mu') = \hat{\delta}(\mu')D(E'/\hat{E}(\mu')),
    \label{eq:fp_see_yield}
\end{equation}
\begin{equation}
	\hat{\delta}(\mu') = \hat{\delta}_{ts}[1 + t_1(1 - \mu'^{t_2})],
\end{equation}
\begin{equation}
	\hat{E} = \hat{E}_{ts}[1 + t_3(1 - \mu'^{t_4})],
\end{equation}
\begin{equation}
	D(x) = \frac{sx}{s - 1 + x^s},
\end{equation}
where $\hat{\delta}_{ts}$, $\hat{E}_{ts}$, $s$, $t_1$, $t_2$, $t_3$, and $t_4$ are fitting parameters.

The emission spectrum for the true secondary electrons is expressed by
\begin{widetext}
\begin{equation}
    \begin{split}
        \frac{\partial\delta_{ts}}{\partial E}\bigg(E,E',\mu'\bigg) = \sum_{n = 1}^M\frac{nP_{n,ts}(E', \mu')(E/\varepsilon_n)^{p_n - 1}\exp{(-E/\varepsilon_n)}}{\varepsilon_n\Gamma(p_n)P(np_n,E'/\varepsilon_n)}P\big((n-1)p_n, (E' - E)\varepsilon_n\big),
    \end{split}
    \label{eq:fp_see_spectrum}
\end{equation}
\begin{equation}
        P_{n,ts}(E', \mu') = \binom{M}{n}\Bigg(\frac{\delta_{ts}(E', \mu')}{M}\Bigg)^n\Bigg(1 - \frac{\delta_{ts}(E', \mu')}{M}\Bigg)^{M - n},
\end{equation}
\end{widetext}

where $p_n$, $\epsilon_n$, are fitting parameters determined from beam data. $\Gamma(\cdot)$ is the delta function, and $P(\cdot,\cdot)$ is the normalized incomplete delta function. The summation is of the probability $n$ secondary electrons up to a total of $M$ being emitted. $M$ theoretically goes to infinity, but $M=10$ is sufficient for high accuracy.\par
\begin{figure*}
    \includegraphics[width=0.49\linewidth]{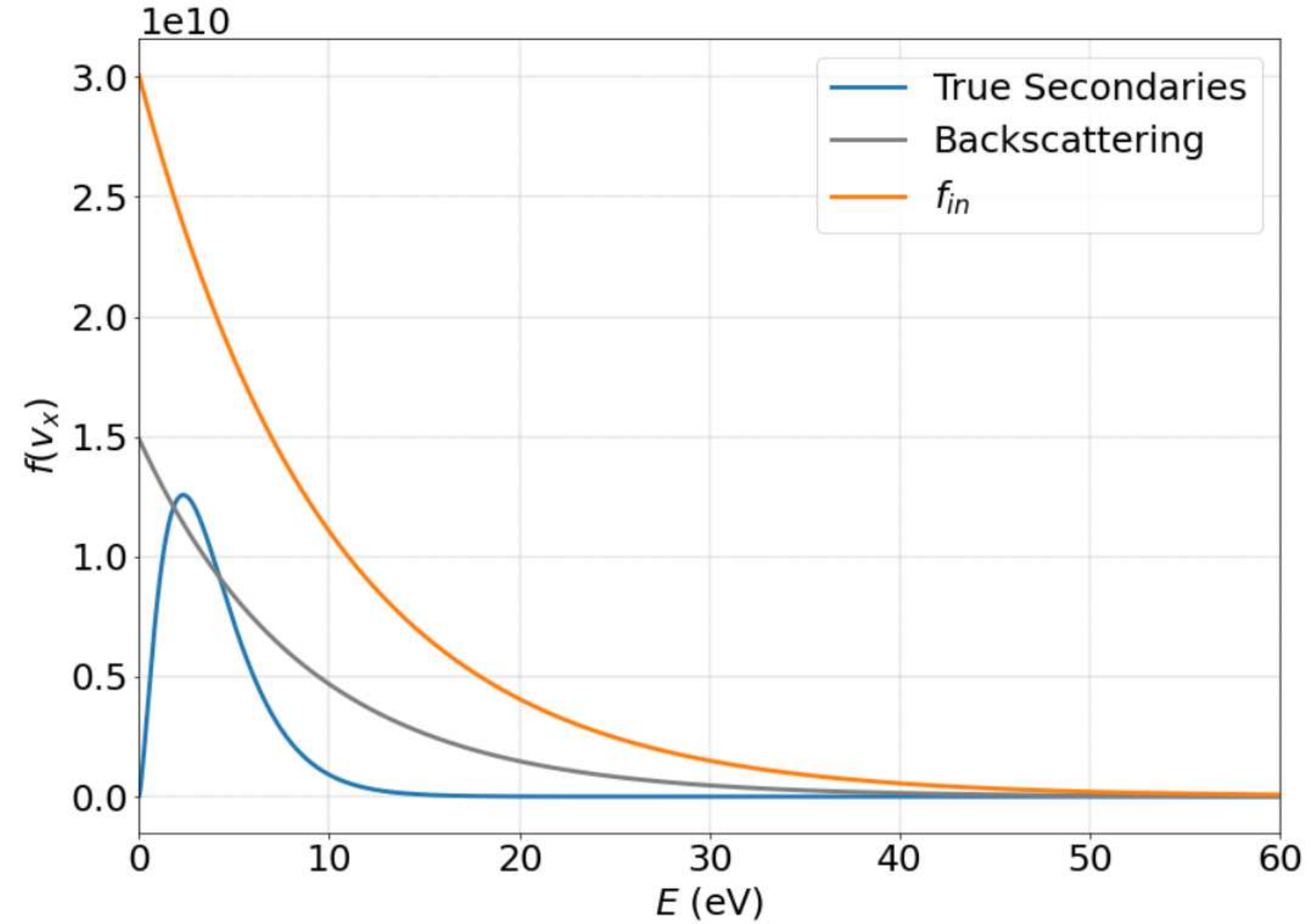}
    \includegraphics[width=0.49\linewidth]{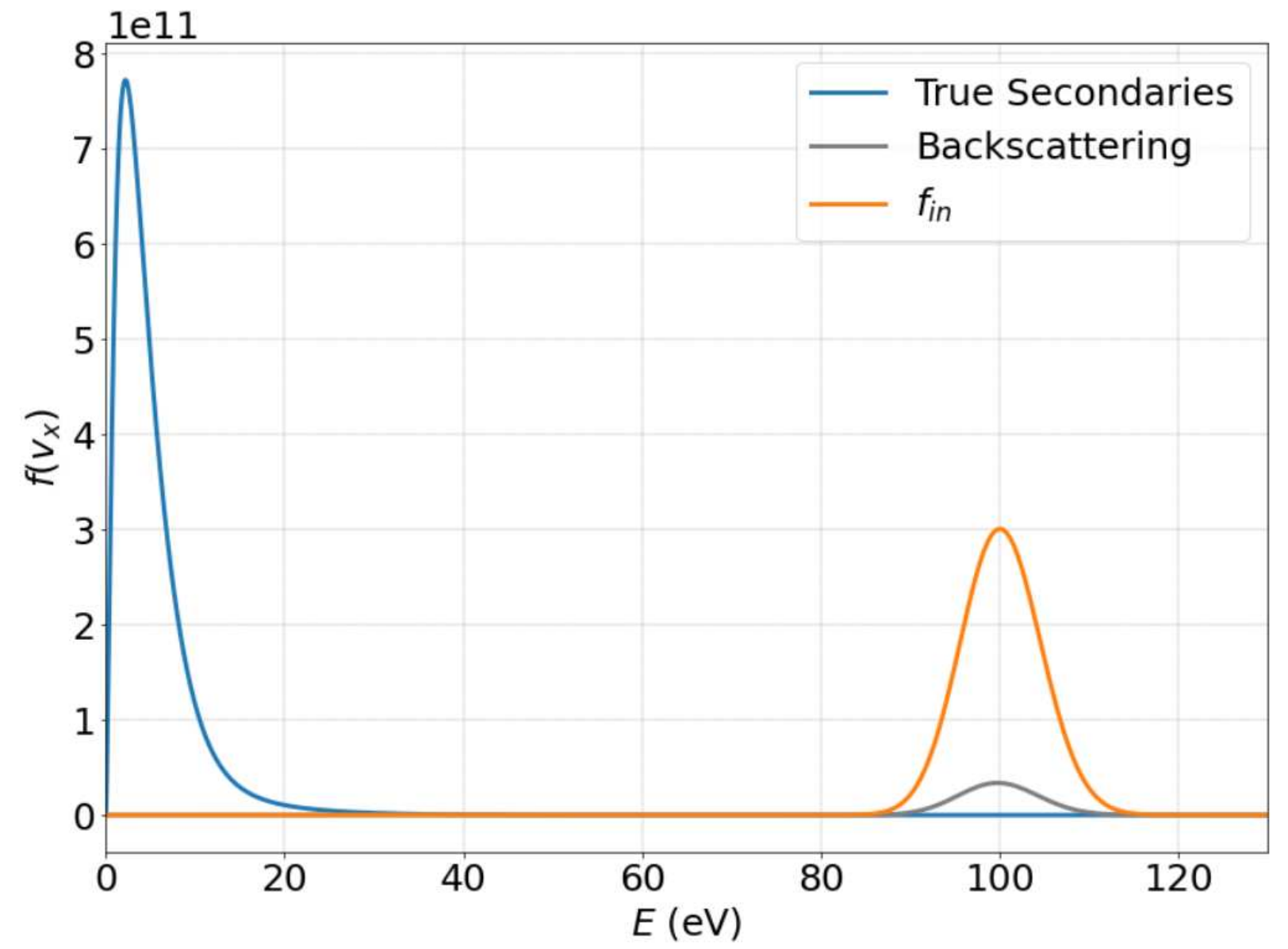}
    \caption{\label{fig:energy_spectra} Plot with the elastic and inelastic emission curves in the low (left) and high (right) energy regimes. At low energy, backscattering contributes significantly and the total emission is concentrated at low energy where the incoming distribution is. For the high energy beam, the true secondary distribution dominates emission but there is a small elastic peak at the beam energy.}
\end{figure*}
There can be a fair amount of variance in the emission data curves for materials depending on methodology and treatment of the material prior to testing \cite{baglin2001,larciprete2013,cimino2015}; for simplicity's sake, here we use the copper parameters calculated by Furman \& Pivi in Table 1 of \cite{furman2002}, used for the fits in Fig.~\ref{fig:fp_fit}. Fits of this same model can be done to any desired dataset, however.\par
The Furman-Pivi model describes the emission of a monoenergetic beam. To apply this to a continuous impacting distribution, simplifications must be made. We can obtain the cumulative emission curve of an incoming distribution by generating a unique emission spectrum for a wide range of incoming points $\mathbf{v}'$ along the distribution and summing these emission spectra to get the total. The individual emission spectrum can be determined from $\frac{\partial\delta}{\partial E}$ by using the definition of yield as the ratio of fluxes.\par
This relation can be written as
\begin{equation}
    \begin{split}
        \int_{\mathcal{V}^-_{\beta}}\mu(\mathbf{v})\frac{\partial\delta}{\partial E}\bigg(E(\mathbf{v}), E'(\mathbf{v}'), \mu'(\mathbf{v})\bigg)\mathcal{J}_{E,\mu}(\mathbf{v})d\mathbf{v}\\
        = \frac{\int_{\mathcal{V}_{\beta}^-} \mathbf{n}\cdot\mathbf{v}f_{ts}(\mathbf{v})d\mathbf{v}}{\mathbf{n}\cdot\mathbf{v}'f_{in}(\mathbf{x}_{wall},\mathbf{v}')d\mathbf{v}},
    \end{split}
    \label{eq:fp_flux_ratio}
\end{equation}\par
where $\mathcal{J}_{E, \mu}(\mathbf{v}) = \left|\frac{\partial(E, \mu)}{\partial(\mathbf{v})}\right|$ is the Jacobian matrix for the coordinate transformation. This simplifies in 1V for $v_x$ to give us the emission spectrum for each incoming monoenergetic beam,
\begin{equation}
    f_{ts}(v_x) = \frac{m_e}{q_0}\frac{\partial\delta}{\partial E}\bigg(E(v_x), E'(v'_x), \mu'=1\bigg)v'_xf_{in}(x_{wall}, v'_x)dv_x.
    \label{eq:monoenergetic_beams}
\end{equation}
Fig.~\ref{fig:energy_spectra} demonstrates the application of this calculation to two distributions, one a high energy $\SI{100}{eV}$ beam, the other a cold Maxwellian with a temperature of $\SI{10}{eV}$. Also included in the figures are the backscattered spectra, which will be discussed in greater depth in Section IV-C. There are two primary features we would draw attention to. First, the location and width of the secondary electron emission distribution do not change significantly based on the incoming energy range, only the magnitude of the emitted spectrum. Second, we would note that even for the cold distribution where we expect emission of true secondaries to be low, the high energy tail is sufficiently populated to create a significant emitted distribution.\par
Instead of implementing the Furman-Pivi spectrum directly into the software, as it is mathematically complicated and computationally expensive to compute during run time, we instead will substitute one of the $f_{ts}(\mathbf{v})$ equations from the previous section and use the full Furman-Pivi only for purposes of comparing model accuracy. As the Maxwellian is less accurate to observed emission data, we will only look at the Gaussian and Chung-Everhart models. The key step here is that Eq.~\ref{eq:fp_see_yield}, the energy-dependent Furman-Pivi SEY equation, is substituted in for $\delta_{ts}$ in the calculation of the normalization factors from the solutions of Eq.~\ref{eq:norm_calc}, ensuring that the total yield remains the same (``conserved" during the spectrum substitution) despite the simpler spectrum model.\par
\begin{figure}
    \includegraphics[width=1.0\linewidth]{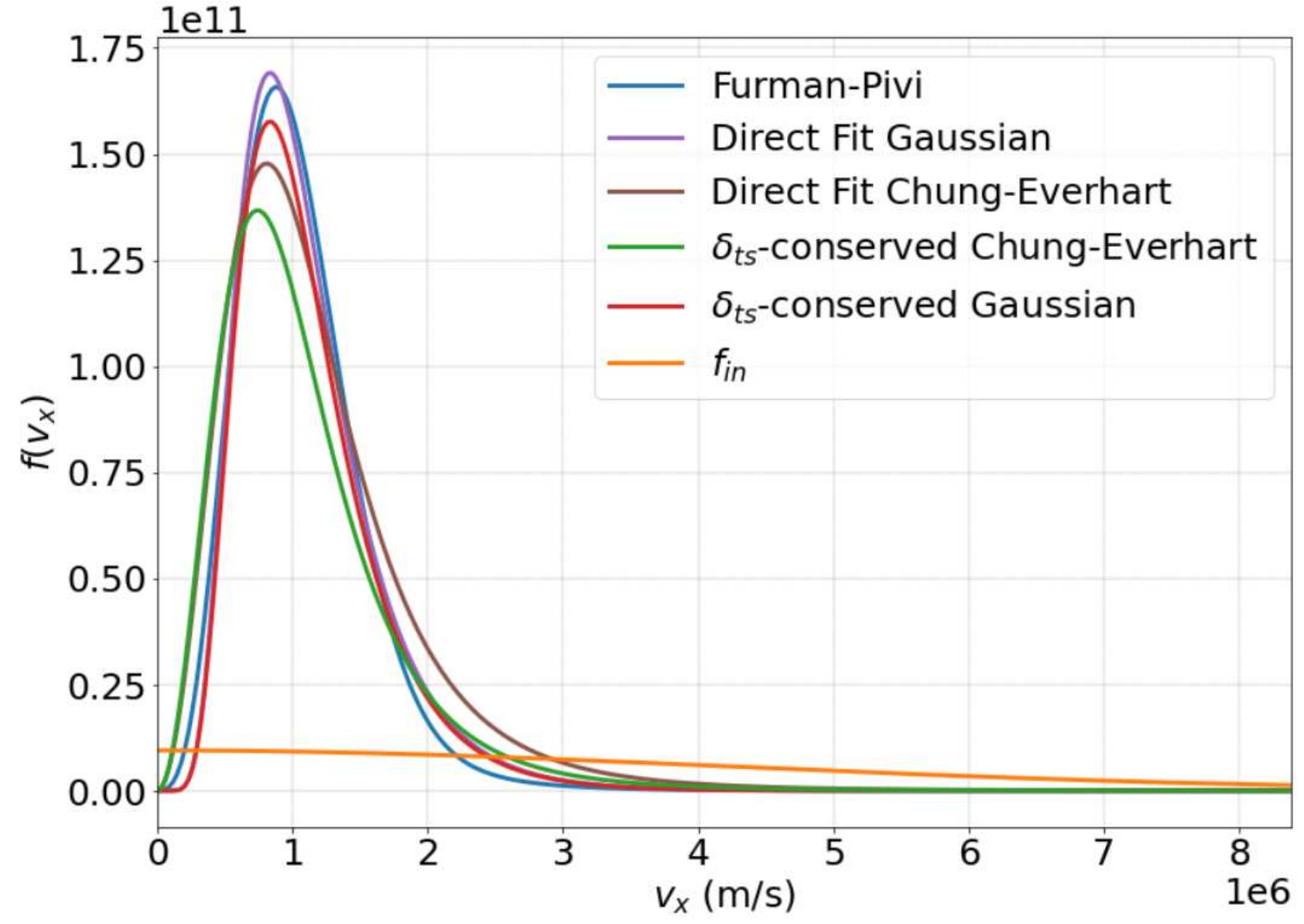}
    \caption{\label{fig:see_spectra} Comparison of several different options for the cumulative true secondary emission spectrum of a $T_e = \SI{100}{eV}$, $n_e = \SI{1.0e17}{m^{-3}}$ Maxwellian. Direct fits match the theoretical Furman-Pivi spectrum quite well, but the overestimation of the high velocity tail is consequential when looking at the total yield (see Fig.~\ref{fig:see_yield}).}
\end{figure}

\begin{figure}
    \includegraphics[width=1.0\linewidth]{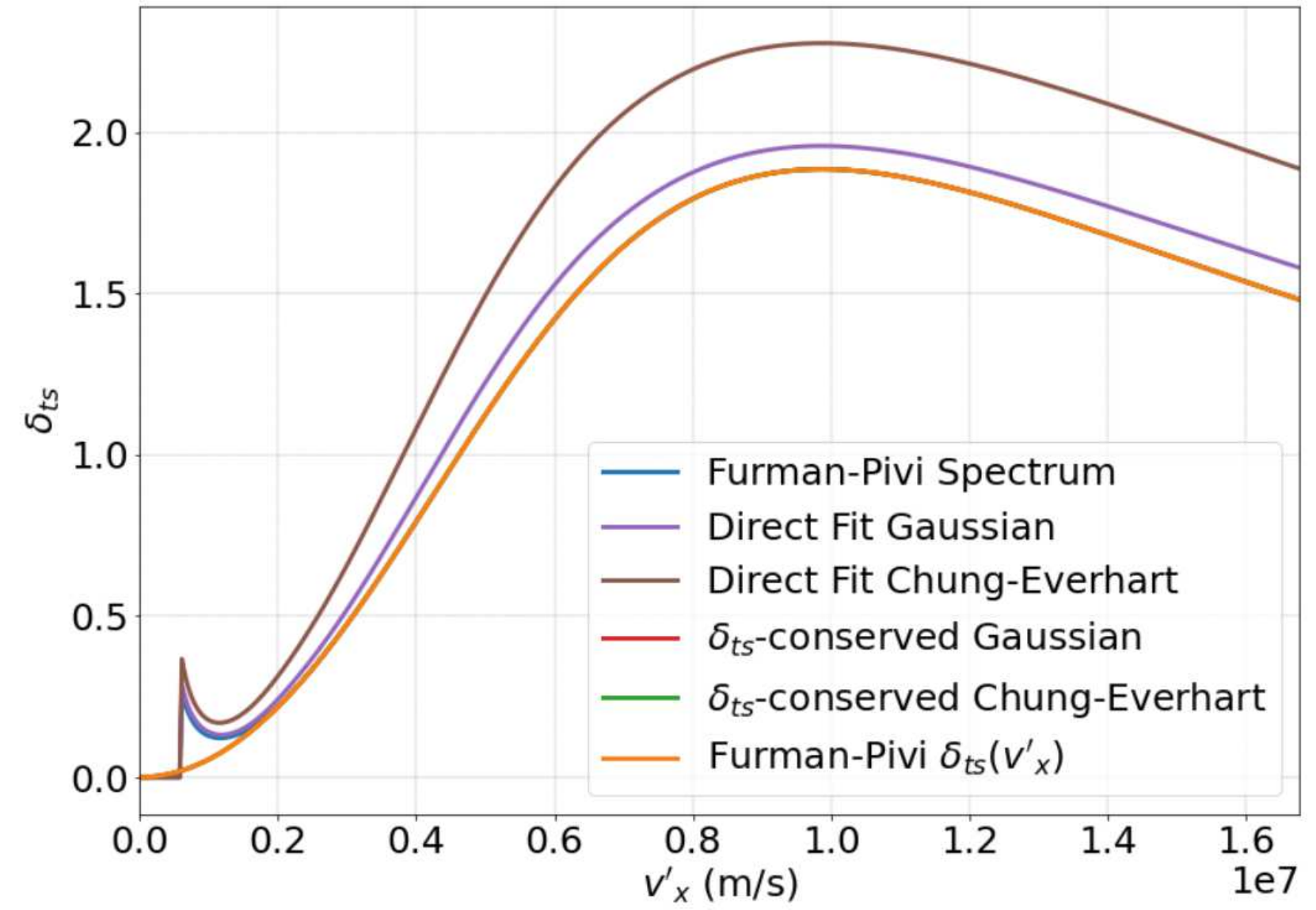}
    \caption{\label{fig:see_yield} Comparison of the Furman-Pivi yield curve to total yield obtained by integrating to get the emission spectrum flux for a range of incoming energies. The $\delta$-conserving models are, as expected, exact to the Furman-Pivi yield curve. The Furman-Pivi spectrum and fits directly to it diverge at very low energy from the theory, as the Furman-Pivi fitting parameters are obtained from high energy beam data. Fits directly to Furman-Pivi overshoot the total yield.}
\end{figure}
Fig.~\ref{fig:see_spectra} shows these spectra for an incoming Maxwellian with an electron temperature of $\SI{100}{eV}$. In addition to the $\delta_{ts}$-conserving models, direct fits of the Gaussian and Chung-Everhart to the Furman-Pivi distribution are shown for comparison. While these direct fits are closer in terms of overall error in the distribution from the Furman-Pivi curve, the excess in the higher velocity region causes them to considerably overshoot the total yield, as shown in Fig.~\ref{fig:see_yield}. The SEY curves in Fig.~\ref{fig:see_yield} are obtained by integrating over each unique emission spectrum produced by an incoming $v_x'$ point to get the outgoing flux, and taking the ratio with the incoming flux $v'_xf_{in}(v'_x)dv_x$. When compared to Eq.~\ref{eq:fp_see_yield}, the yield conserving cases do, as expected, match perfectly to the theoretical SEY curve. Curiously, at very low energies the theoretical Furman-Pivi emission spectrum does not, and indeed must be cut off to avoid a sharp increase near zero. This is not actually so surprising given that the Furman-Pivi spectrum fitting parameters are obtained from high energy data. These work extremely well for the majority of the energy range, but there is a particular dearth of emission data at very low energies, and in this region they begin to break down. As the direct fit curves are done to the theoretical Furman-Pivi spectrum, they share this quirk. It should be noted that since yield is based on flux, not density, low energy particles only contribute a minor amount compared to the high-velocity tail of the distribution, so this divergence is not particularly bothersome in any case.\par
Overall, the Gaussian provides a better match to the emission spectrum, but the Chung-Everhart bears a distinct advantage. It's sole fitting parameter, $\phi$, is the material work function, and thus can be determined adequately without any emission spectrum data. The Gaussian, on the other hand, requires estimates of $E_0$ and $\tau$. $E_0$ represents the peak location, which can in a pinch be approximated per Chung \& Everhart as $\phi/3$ \cite{chung1974}. However, $\tau$ can only be estimated by some fit to existing spectrum data, which can be quite rare and not readily available for some materials.\par
In either case, however, adopting this simple modelling method of swapping out the complicated spectrum fit while relying on the fits to SEY data gives us fairly accurate representations of the emitted spectra which perfectly conserve the total yield.

\subsection{Elastic Model}

For elastic emission $E = E'$, $\mu = \mu'$. Thus, this boundary condition is simply the direct scaling of the incoming distribution by some $\delta_e$,
\begin{equation}
    f_e(\mathbf{v}) = \delta_e(\mathbf{v})f_{in}(\mathbf{x}_{wall}, \mathbf{v}').
\end{equation}

There are a number of available choices for the yield function. The one used in this work is the term derived as part of the Furman-Pivi model
\begin{equation}
    \begin{split}
        \delta_{e0}(E') = P_{1,e}(\infty) + [\hat{P}_{1,e} - P_{1,e}(\infty)]\\
        \exp{\big[(-|E' - \hat{E}_e|/W)^p/p\big]},
    \end{split}
\end{equation}
where $P_{1,e}(\infty)$, $\hat{P}_{1,e}$, $\hat{E}_e$, $W$, and $p$ are fitting parameters. This is plotted next to the true secondary curve in Fig.~\ref{fig:energy_spectra}. Being elastic, the backscattered population is concentrated at the same location as the incoming distribution regardless of energy regime. The magnitude of the emitted distribution, however, is much greater in the low energy regime as would be expected from Fig.~\ref{fig:fp_fit}.\par
More complete low-temperature emission measurements by Cimino et al \cite{cimino2004,larciprete2013,cimino2015} suggest that the reflection function may actually go to unity at $E'=0$ and present an alternate model based on the quantum mechanical probability for the reflection of a plane-wave that fits well to their data. For dielectric materials, the work of Bronold and Fehske gives accurate fits to reflection and rediffusion data at low energies \cite{bronold2015}; numerical implementation of this model has been described by Cagas et al \cite{cagas2020}.

\begin{figure}
    \includegraphics[width=1.0\linewidth]{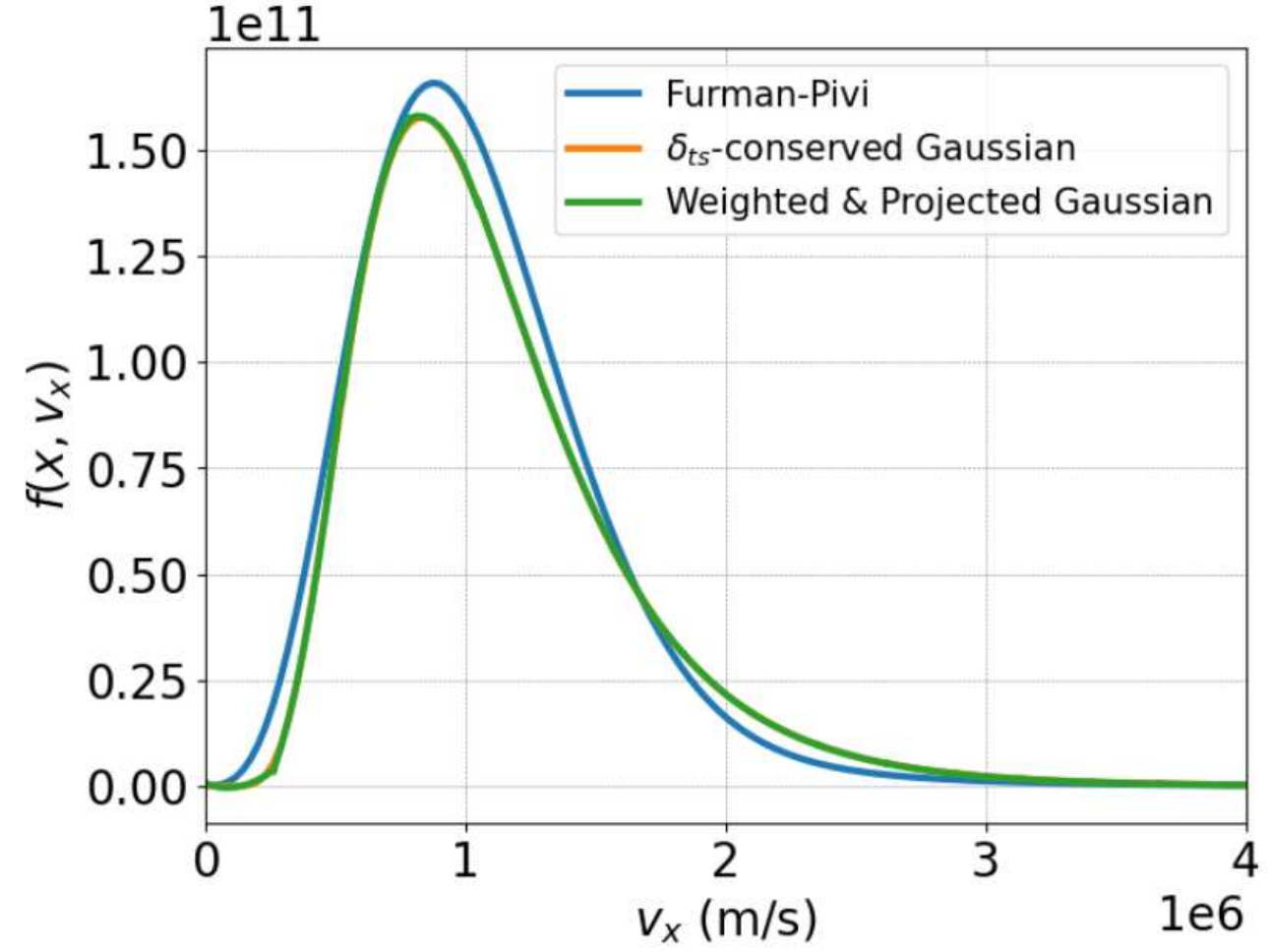}
    \caption{\label{fig:projection_comparison} Comparison of exact Furman-Pivi spectra summed across incoming cells, yield-conserving Gaussian, and simplified Gaussian projected onto the basis. Loss of accuracy is minimal despite the simplifications.}
\end{figure}

\section{Discrete Implementation}

\begin{figure*}
    \includegraphics[width=0.49\linewidth]{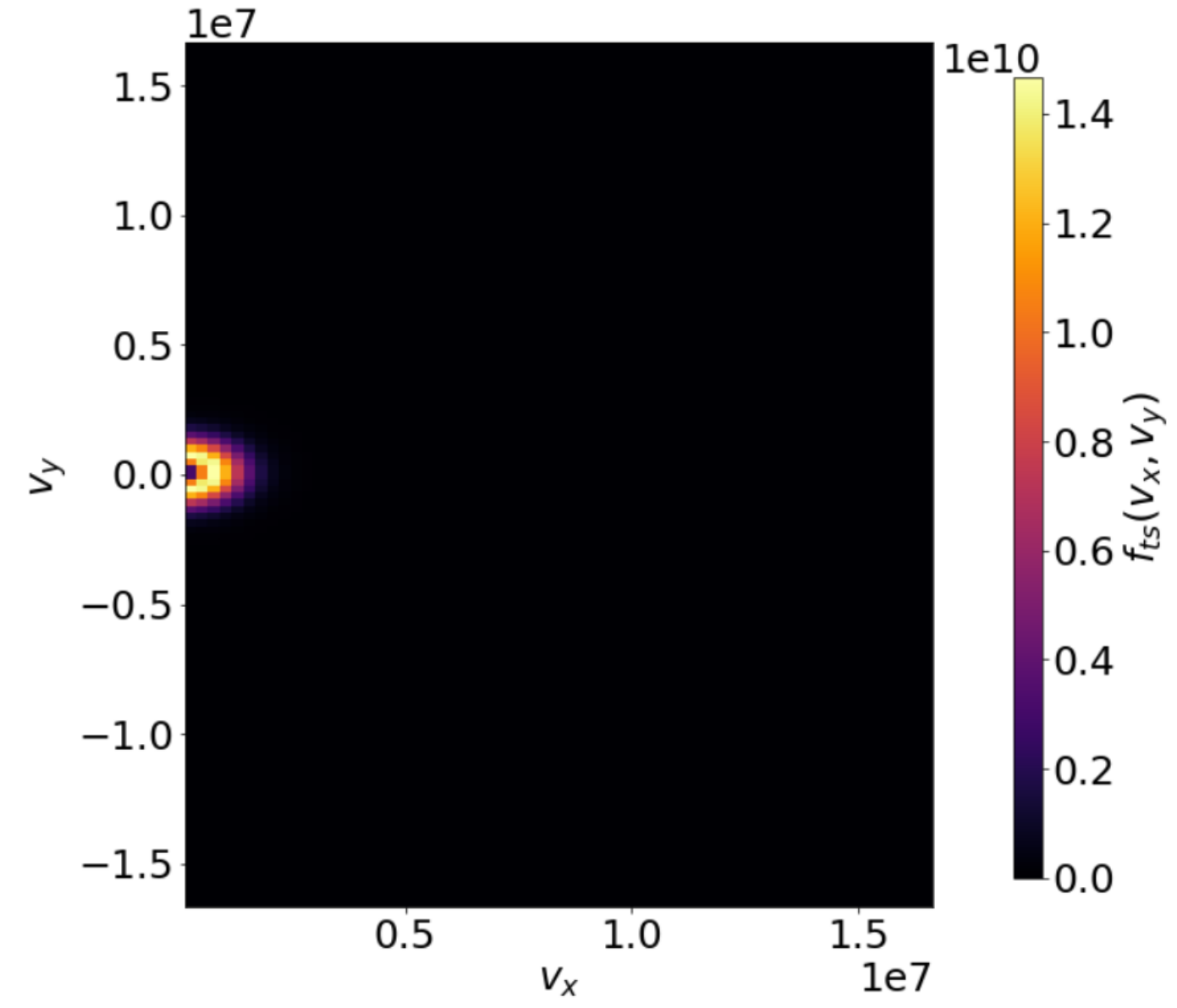}
    \includegraphics[width=0.49\linewidth]{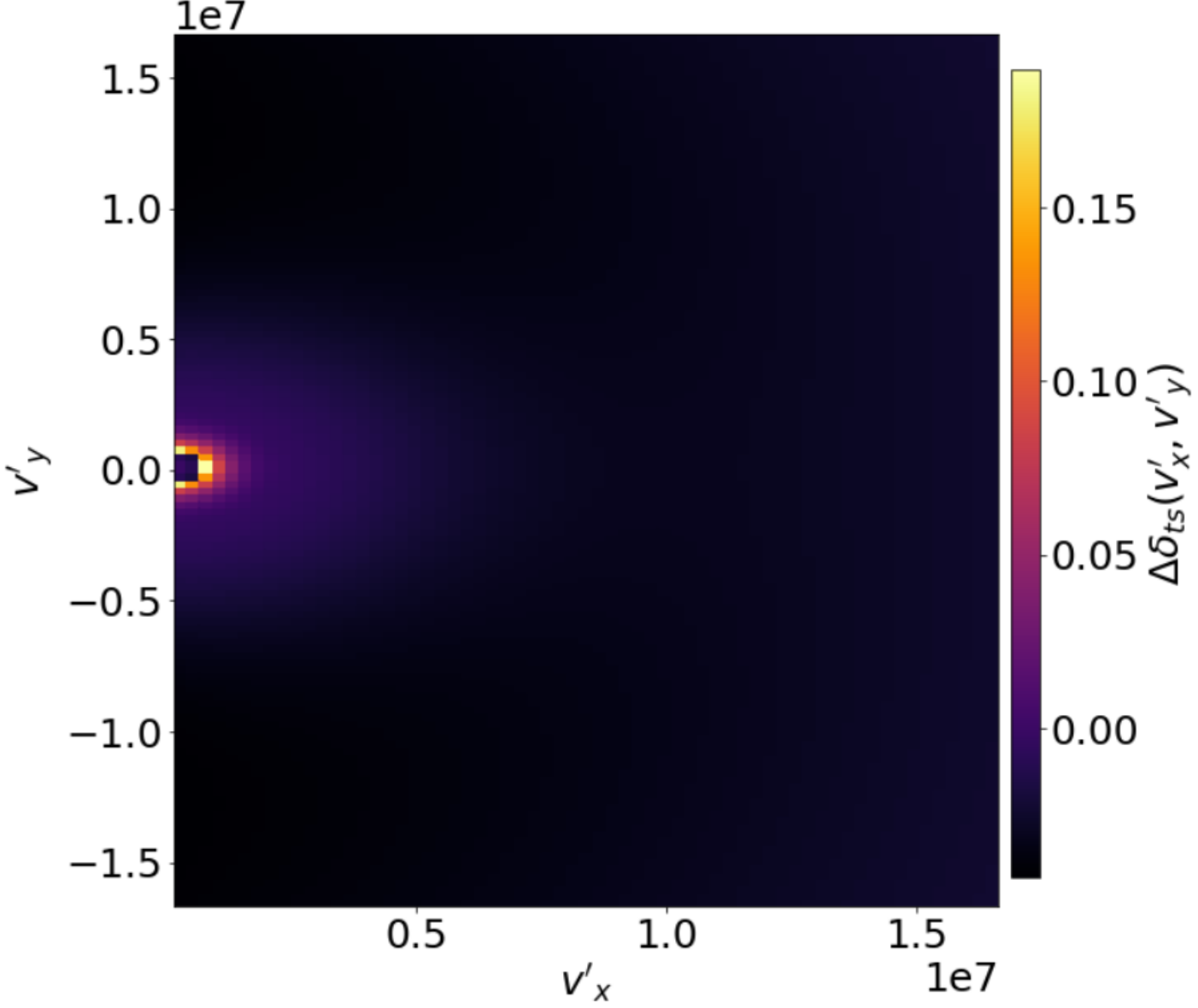}
    \includegraphics[width=0.49\linewidth]{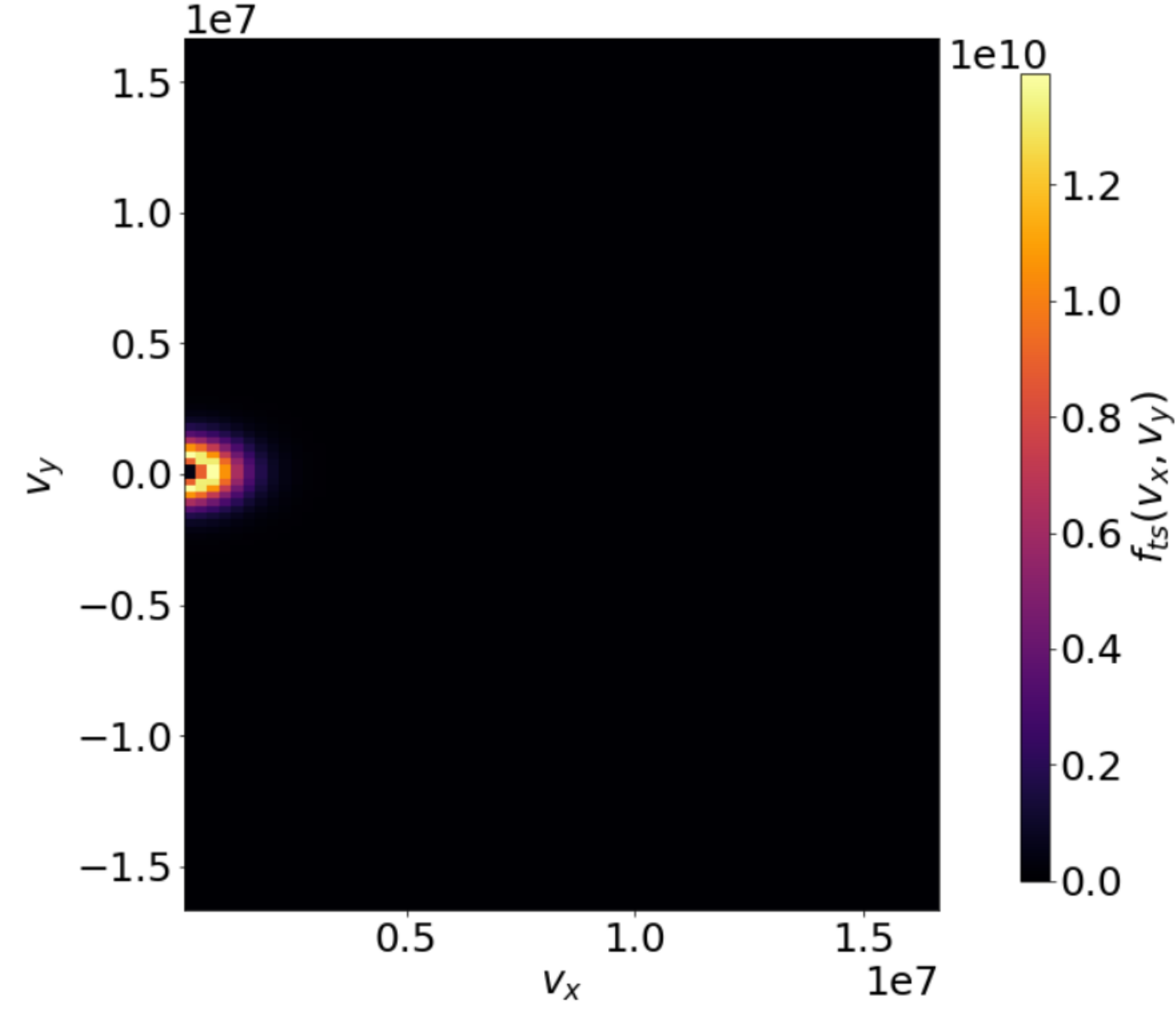}
    \includegraphics[width=0.49\linewidth]{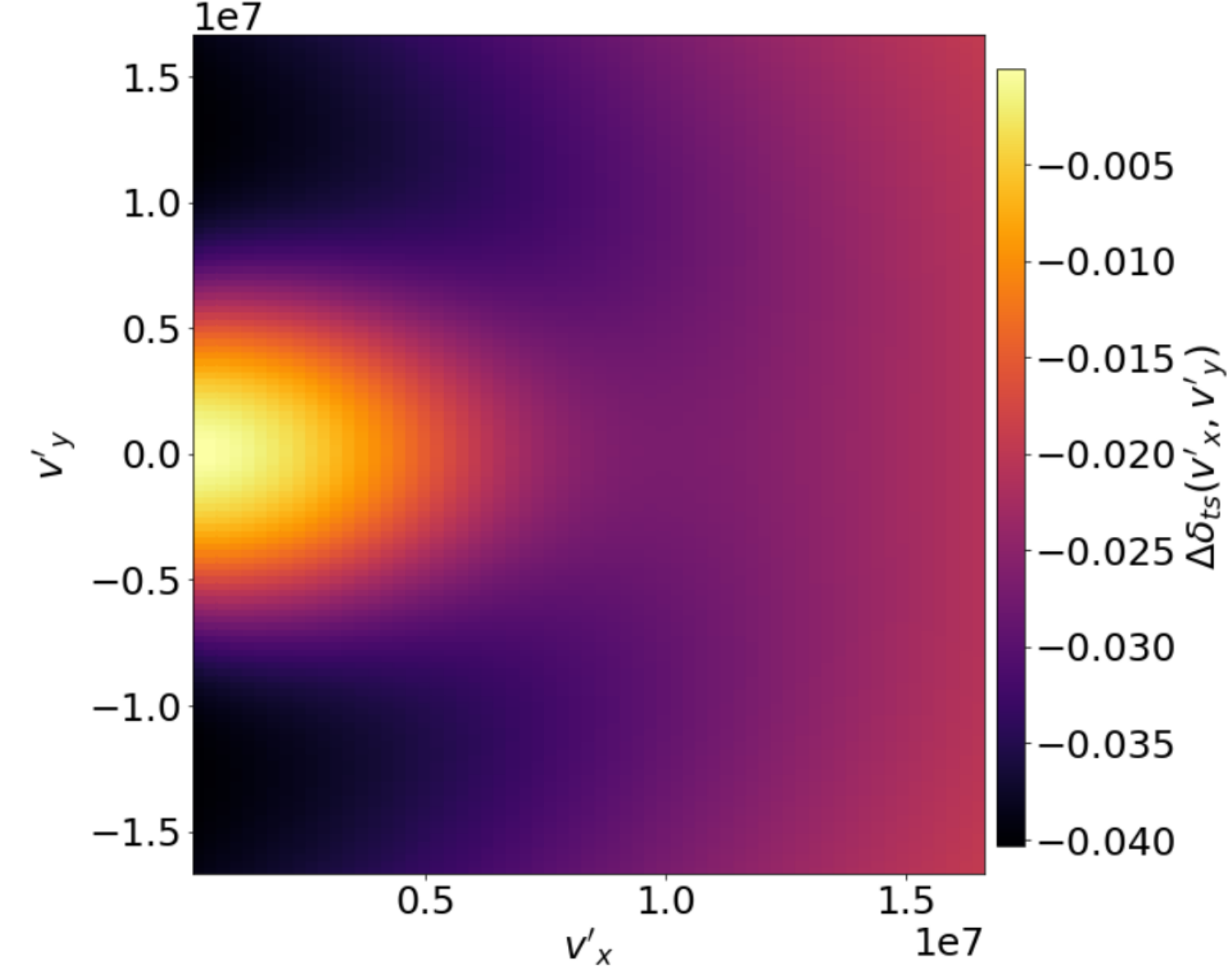}
    \caption{\label{fig:2v_spectrum} 2V extension of the emission model for the Furman-Pivi theory (top) and Gaussian (bottom). Despite the angular dependence of $\frac{\partial\delta_{ts}}{\partial E}$, the two spectrum distributions are azimuthally uniform (left). Integrating over the Furman-Pivi spectrum to get the yield results in the same error spike at low energy seen in 1V, while the Gaussian has low error throughout the domain (right).}
\end{figure*}

\subsection{Emission Spectrum}

The implementation of these models on the discrete level requires some modifications. The previous plots were generated from the summation of unique spectra from hundreds of incoming monoenergetic beams along the impacting distribution. Thus, an immediate necessary simplification is to cut the number of emission spectra down to one for each cell center velocity. Requiring even each cell to store its own unique emission spectrum can be impractical, however, particularly for multidimensional simulations on large grids. Thus, it is very desirable to consolidate into a single emission spectrum. We take advantage of the fact that cell emission spectra share a shape, and the observation made earlier that the fitting parameters are not significantly different across the domain for different incoming energies; the major difference lies only in the normalization factor. Therefore, scaling the emitted distribution by the cumulative SEY of the entire incoming distribution using a weighted average should suffice. The weighted average is done using a loop over the velocity space skin cells, the cell center fluxes and the 0th expansion coefficient of the distribution function,

\begin{equation}
    \bar{\delta}_{ts} = \frac{\sum_{j=1}^{N_v}v_c^{j}\hat{f}^j_{\alpha,0}\delta_{ts}(v_c^{j})}{\sum_{j=1}^{N_v}v_c^{j}\hat{f}^j_{\alpha,0}}.
    \label{eq:weighted_delta}
\end{equation}

This effectively weights each value of $\delta$ by its contribution to the total incoming flux. Then, we simply substitute in this value for $\delta_{ts}$ in the calculation for $C$, generating a single, overall normalization factor. The chosen emission spectrum function $f_{ts}$ is projected onto the basis in the initialization of the simulation using the methods described in Section II-B, and scaled each time step by the freshly calculated $C$.\par
The calculation of $C$ requires us to obtain the flux into the wall, $\Gamma_{\alpha}^+$. This flux is most accurately described as the distribution lost to the wall during a time step, or, in numerical terms, the RHS terms of Eq.~\ref{eq:vlasov} in the ghost cell, $\frac{\partial f_{h\alpha}^g}{\partial t}$. Taking the integrated density of this quantity gives us the total flux,

\begin{equation}
    \begin{split}
        \Gamma_{\alpha}^+ = \int_{\mathcal{I}}\frac{\partial f^{g+}_{h\alpha}}{\partial t}(\boldsymbol{\eta}_{\mathbf{x}}, \boldsymbol{\eta}_{\mathbf{v}})\mathcal{J}_{\mathbf{x},\mathbf{v}}(\boldsymbol{\eta}_{\mathbf{x}}, \boldsymbol{\eta}_{\mathbf{v}})d\boldsymbol{\eta}_{\mathbf{x}}d\boldsymbol{\eta}_{\mathbf{v}},
    \end{split}
\end{equation}

where $\mathcal{J}_{\mathbf{x},\mathbf{v}}(\boldsymbol{\eta}_{\mathbf{x}}, \boldsymbol{\eta}_{\mathbf{v}}) = \left|\frac{\partial(\mathbf{x}, \mathbf{v})}{\partial(\boldsymbol{\eta}_{\mathbf{x}}, \boldsymbol{\eta}_{\mathbf{v}})}\right|$ is the Jacobian matrix transformation from physical to logical space.\par
The combined steps for calculating the boundary condition are described in Alg.~\ref{algo:bc_calc}, which is inserted into the RK update (see Alg.~\ref{algo:rk_step}). Each update step after the RHS terms are calculated but before the forward Euler is advanced, $\frac{\partial f}{\partial t}$ in the ghost cell is used to calculate the distribution lost to the wall over the course of a time step (i.e. the flux into the wall). A loop is done over the skin cells to calculate the weighted yield as described by Eq.~\ref{eq:weighted_delta}, which is used to calculate the normalization factor. After the time step is advanced, the elastic portion is calculated from the updated skin cell, and both this distribution and the scaled emission spectrum are applied to the ghost cells.\par
The resulting function is compared to the theoretical Furman-Pivi spectrum, and the yield conserving Gaussian spectrum in Fig.~\ref{fig:projection_comparison}. Even with this averaging and discretization, the resultant distribution suffers only minor loss of accuracy.\par
The methodology outlined here can be extended beyond electron-impact SEE, and the algorithm allows for multiple impact species to be considered. It merely requires an equivalent model to replace the Furman-Pivi model aimed towards ion-impact SEE or whichever desired emission phenomenon.

\subsection{Elastic Emission}

As stated, the elastic boundary condition is the direct scaling of the incoming distribution in the skin cells by some $\delta_e$, which is then reflected to become the outgoing distribution in the ghost cells. In the discrete sense, this becomes:
\begin{equation}
    f^{g-}_{h\beta}(\mathbf{x}, \mathbf{v}) = \delta_{he}(\mathbf{v})f^{s+}_{h\beta}(\mathbf{x}^-, \mathbf{v}^-).
\end{equation}
Here, the coordinates $\mathbf{x}^-$, $\mathbf{v}^-$ denote the incoming coordinates being transformed into their outgoing coordinates\footnote{This entails a flipping of the sign for the boundary coordinates. For example, at the $x$-boundary $\mathbf{x}^-, \mathbf{v}^- = (-x, y, z), (-v_x, v_y, v_z)$.}.
In the simplest cases, as touched on in the section on constant yield, $\delta_e$ may be taken as a constant across velocity space. For implementations where $\delta_e$ is allowed to vary with the incoming energy, it must be implemented as a discrete function. Unlike our implementation of $\delta_{ts}$, this yield is not being used to scale a projected continuous function, and thus cannot be merely evaluated at each cell center and then multiplied by the distribution in that cell. Doing so means multiplying a set of discrete functions by different constant values, which leads to sharper discontinuities in the distribution function at the cell edges and can drive increasing and ultimately fatal numerical noise. Instead, there must be a weak multiplication of the two discrete functions $\delta_{h,e}$ and $f_h$,

\begin{equation}
    \hat{f}^{g-}_{\beta e} = \int_{\mathcal{I}} \delta_{h,e}(\boldsymbol{\eta}_{\mathbf{v}})f^{s+}_{h\beta}(\boldsymbol{\eta}^-_{\mathbf{x}}, \boldsymbol{\eta}^-_{\mathbf{v}})\psi_t(\boldsymbol{\eta}_{\mathbf{x}}, \boldsymbol{\eta}_{\mathbf{v}})d\boldsymbol{\eta}_{\mathbf{x}}d\boldsymbol{\eta}_{\mathbf{v}}.
\end{equation}

\section{Discussion}

We have demonstrated the basic numerical algorithm for implementation of SEE models into a continuum kinetic boundary condition. The result is a fully energy-dependent emission algorithm that robustly handles the full range of incoming energies, and is applicable for any material for which an SEY curve may be obtained. In this section, we will address the question of how introducing angular effects changes the emission spectrum, and conclude by showing the results of implementing this boundary condition in sheath simulations across a couple different parameter regimes.\par

\begin{figure}[htp!]
    \includegraphics[width=1.0\linewidth]{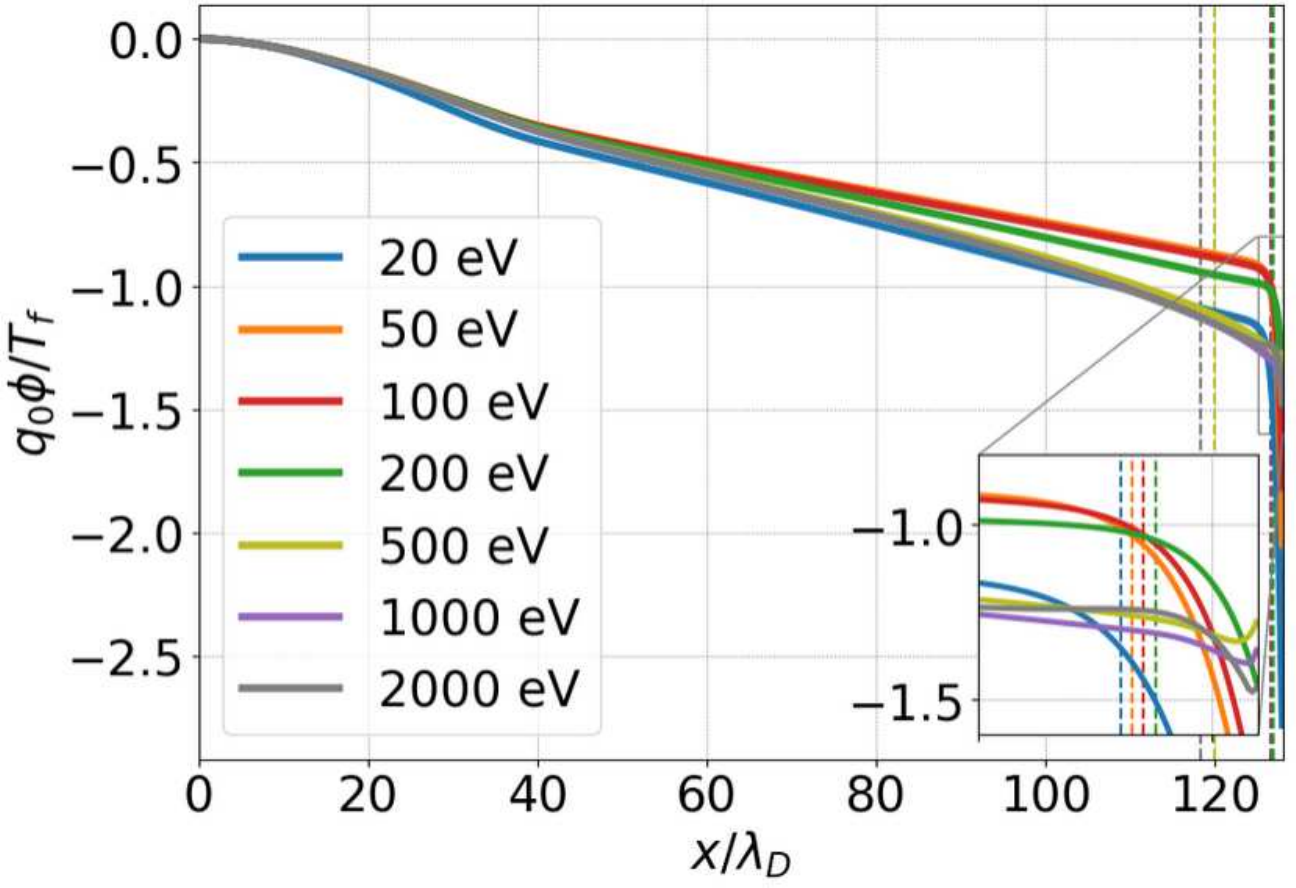}
    \caption{\label{fig:sim_potential} Plot of the potential profiles for the cases at $t\omega_{pe} = 10000$. Dotted lines mark where ions reach the sound speed. The high temperature cases transition to an SCL sheath, while, as expected, the low emission cases remain classical.}
\end{figure}

\subsection{Angular Dependence of the Emitted Distribution}

The plots shown so far have been from emission in a single velocity space dimension. Adding in the second activates an angular dependence in Eq.~\ref{eq:fp_see_yield} and the spectrum. The emission spectrum can be assumed to follow a $\cos{\theta}$ angular distribution \cite{furman2002}. However, this does not translate to merely scaling $f_{ts}$ by a factor of $\mu$. Returning to the flux balance of the emitted distribution and incoming beam, Eq.~\ref{eq:fp_flux_ratio} , when we extend it to 2V for $v_x$, $v_y$, the result is 
\begin{widetext}
\begin{equation}
        f_{ts}(v_x, v_y) = \frac{v'_x}{v_x}\mu(v_x, v_y)f_{in}(x=x_{wall}, v'_x, v'_y)\frac{m_e}{q_0}\frac{\partial\delta}{\partial E}\bigg(E(v_x, v_y), E'(v'_x, v'_y), \mu'(v'_x, v'_y)\bigg)dv_xdv_y.
        \label{eq:2v_flux_ratio}
\end{equation}
\end{widetext}
The solution to the Gaussian spectrum function gained by evaluating Eq.~\ref{eq:norm_calc} in 2V is
\begin{equation}
    C_G = \frac{\delta_{ts}\Gamma_{\alpha}^+m_{\beta}^{3/2}}{4\sqrt{\pi}(q_0E_0)^{3/2}\tau\exp{(9\tau^2/8)}}.
\end{equation}
At first glance this introduces a serious issue for our implementation, as Eq.~\ref{eq:2v_flux_ratio} states that the emitted distribution is dependent on the outgoing velocity, something which makes scaling the uniform $f_{ts}$ by a constant $C$ incorrect. However, the dependence on the emitted normal velocity $v_x$ in Eq.~\ref{eq:2v_flux_ratio} drops out between the numerator of $\mu$ and the ratio $v'_x/v_x$, meaning that while the particles are emitted in a $\cos{\theta}$ distribution, the cumulative emitted distribution function over a time step is actually uniform with angle. Thus, our Gaussian or Chung-Everhart spectrum approximations for $f_{ts}$ do not need to be scaled by any angular dependence, and no modification to the normalization factor is necessary.\par
As indicated, the only angular dependence in the normalization factor is the incoming angle for $\delta_{ts}(E', \mu')$. The Furman-Pivi and Gaussian spectra calculated from Eq.~\ref{eq:2v_flux_ratio} demonstrating this are shown in Fig.~\ref{fig:2v_spectrum}, alongside error plots for the resulting $\delta_{ts}$ gained by integrating these spectra. We see the same error spike that is present in 1V at low energy in the Furman-Pivi plots for the same reason as before. Otherwise, both the Furman-Pivi and Gaussian have very low error in 2V. The same overall algorithm from Alg.~\ref{algo:bc_calc} remains applicable when extended to multiple dimensions.\par

\begin{algorithm}
    \SetAlgoLined
    \tcp{Loop over species, calculation of $\frac{\partial f_{hs}}{\partial t}$ VM-FP RHS terms}
    \tcp{Calculate flux into the wall $\Gamma^+_{\alpha}$ from RHS terms in ghost cell}
    $\Gamma^+_{\alpha} =$ integratedDensity($\frac{\partial f^{g+,n}_{h\alpha}}{\partial t}$);\\
    \tcp{Loop over skin cell phase space into the wall, $\mathcal{X}^s$, $\mathcal{V}^+$, calculate numerator $w_1$ and denominator $w_2$ of weighting function}
    \For{$i\in \mathcal{X}^s$}{\For{$j\in \mathcal{V}_{\alpha}^+$}{$w^i_1 += v_c^{j}\hat{f}^{ij,n}_{\alpha,0}\delta_{ts}(v_c^j)$;\\
    $w^i_2 += v_c^{j}\hat{f}^{ij,n}_{\alpha,0}$;}
    $\bar{\delta}^i_{ts} = w^i_1/w^i_2$;\\
    \tcp{Calculate the normalization factor}
    $C^i =$ calculateNormalizationFactor($\bar{\delta}^i_{ts}$, $\Gamma^{i+}_{\alpha}$);\\}
    \tcp{Advance with Forward Euler and RK update from $f_s^n$ to $f_s^{n+1}$}
    \tcp{Calculation of elastic backscattering spectrum}
    $f_{h,e} =$ multiply($\delta_{h,e}, f^{s,n+1}_{h\beta}$);\\
    \tcp{Accumulate emission spectra to updated distribution ghost cell}
    $f^{g,n+1}_{h\beta} =$ accumulate($Cf_{h,ts},f_{h,e}$);
    
    \caption{Emission spectrum update algorithm} \label{algo:bc_calc}
\end{algorithm}

\begin{figure*}
    \includegraphics[width=0.49\linewidth]{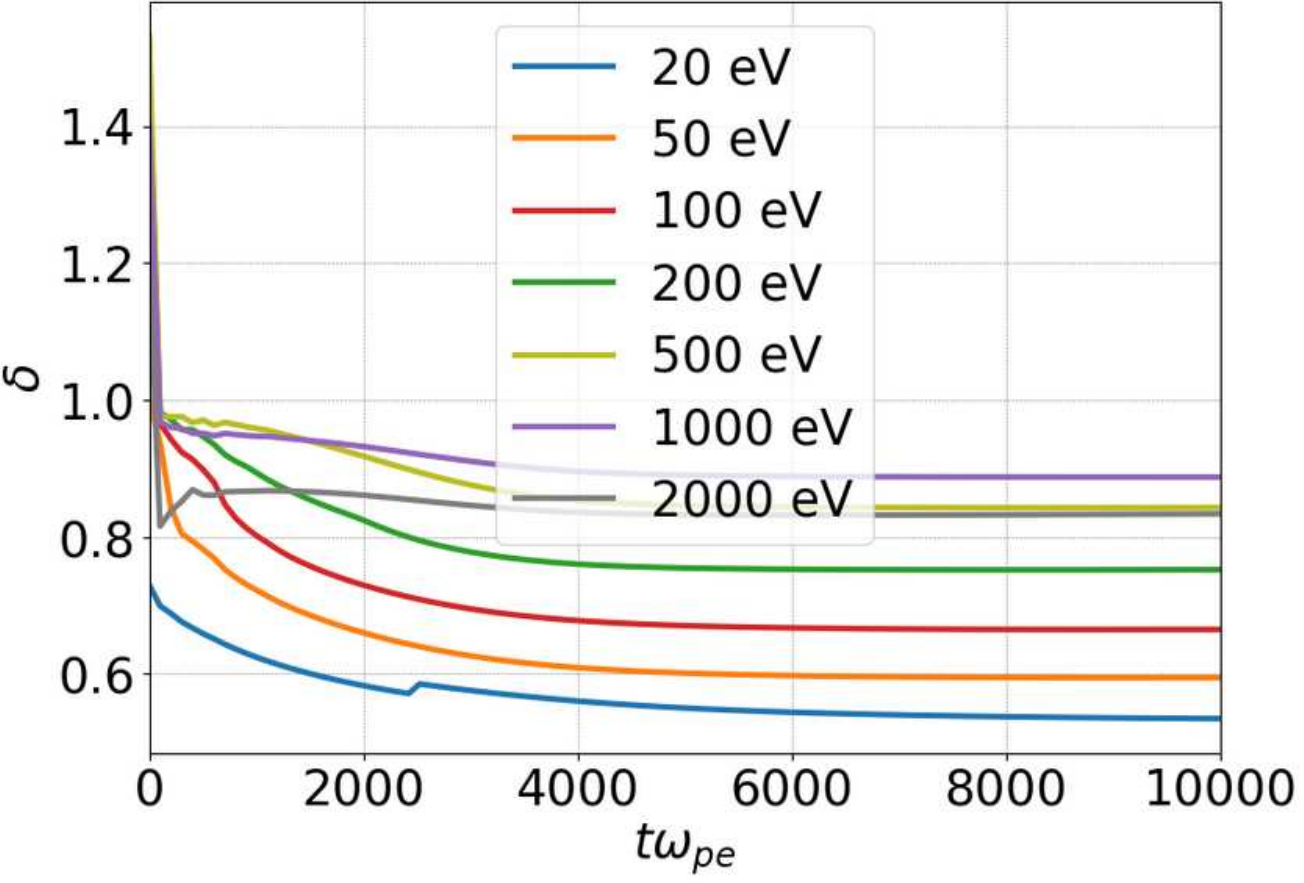}
    \includegraphics[width=0.49\linewidth]{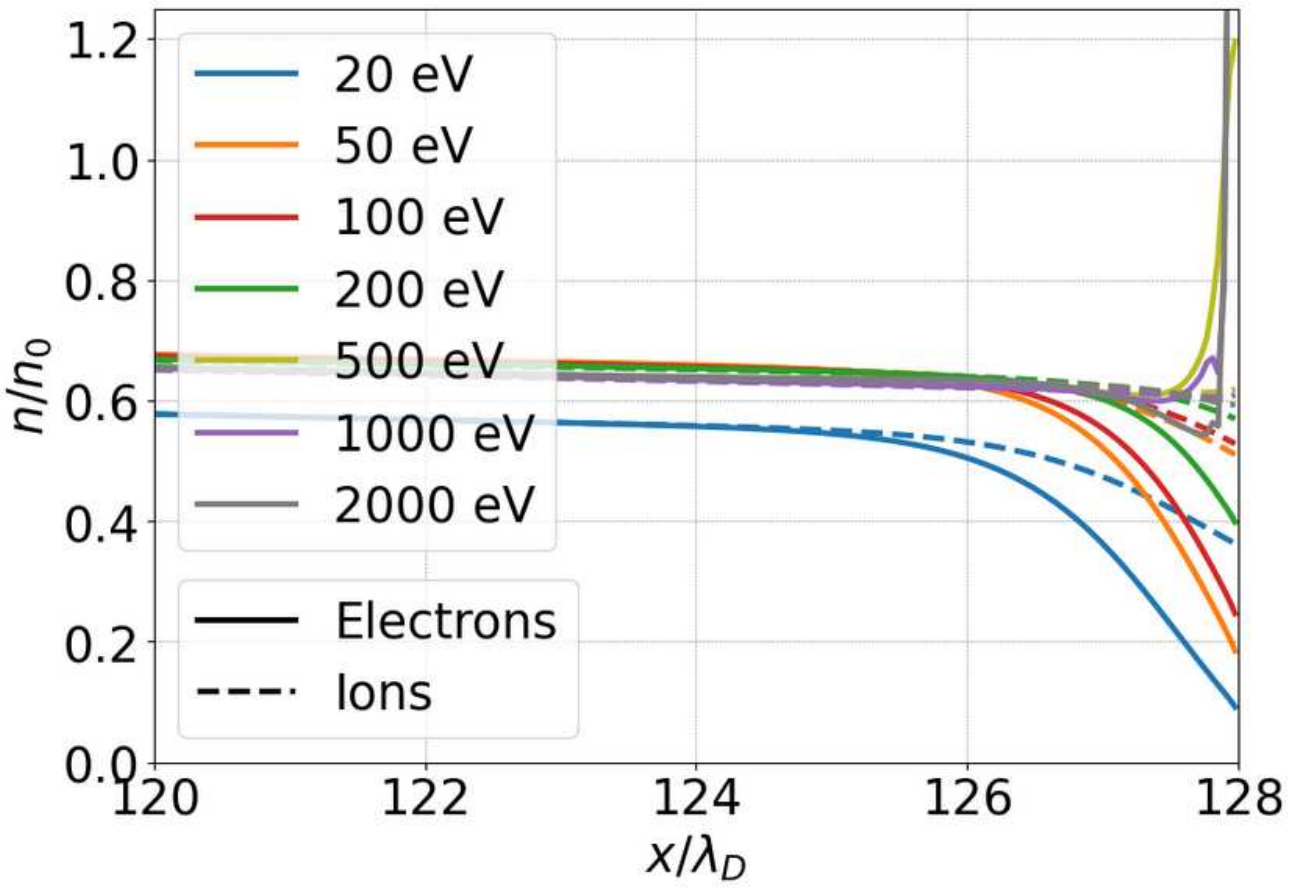}
    \caption{\label{fig:scl_behavior} Time evolution of the emission for different temperature simulations (left) and steady-state density distribution (right). Note that the yield starts at greater than unity for several cases and drops as a result of decompressional cooling during sheath formation.. It continues to be driven lower with time, eventually causing transition back to a classical sheath. The right plot shows the electron and ion density profiles of the cases. Classical sheaths retain a positive space charge, while the SCL cases see heavy accumulation of electrons near the wall.}
\end{figure*}

\begin{figure}
    \includegraphics[width=1.0\linewidth]{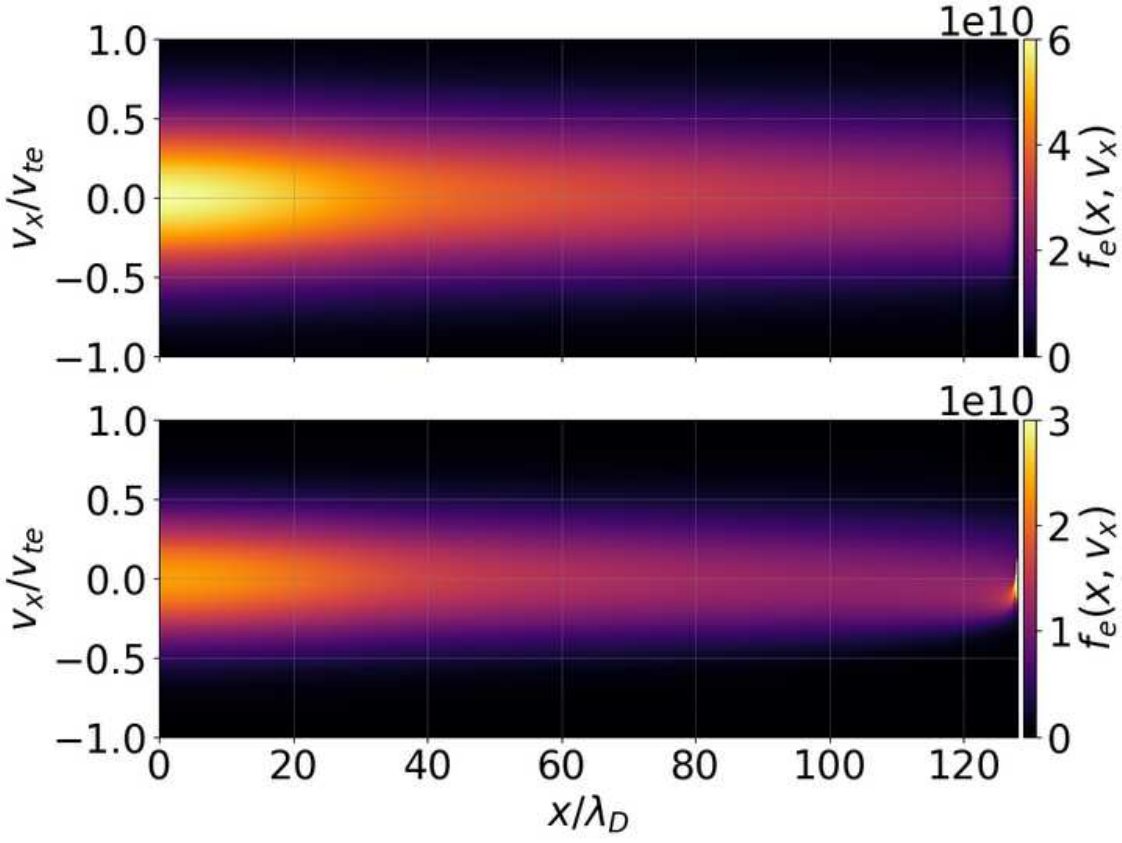}
    \caption{\label{fig:sim_dist} Electron distribution functions for the $\SI{50}{eV}$ (top, classical) and $\SI{500}{eV}$ (bottom, SCL) cases. For the high temperature case with the SCL, a beam of emitted electrons is visible coming off the wall, though it is diffused by collisions soon after.}
\end{figure}

\begin{figure*}
    \includegraphics[width=0.49\linewidth]{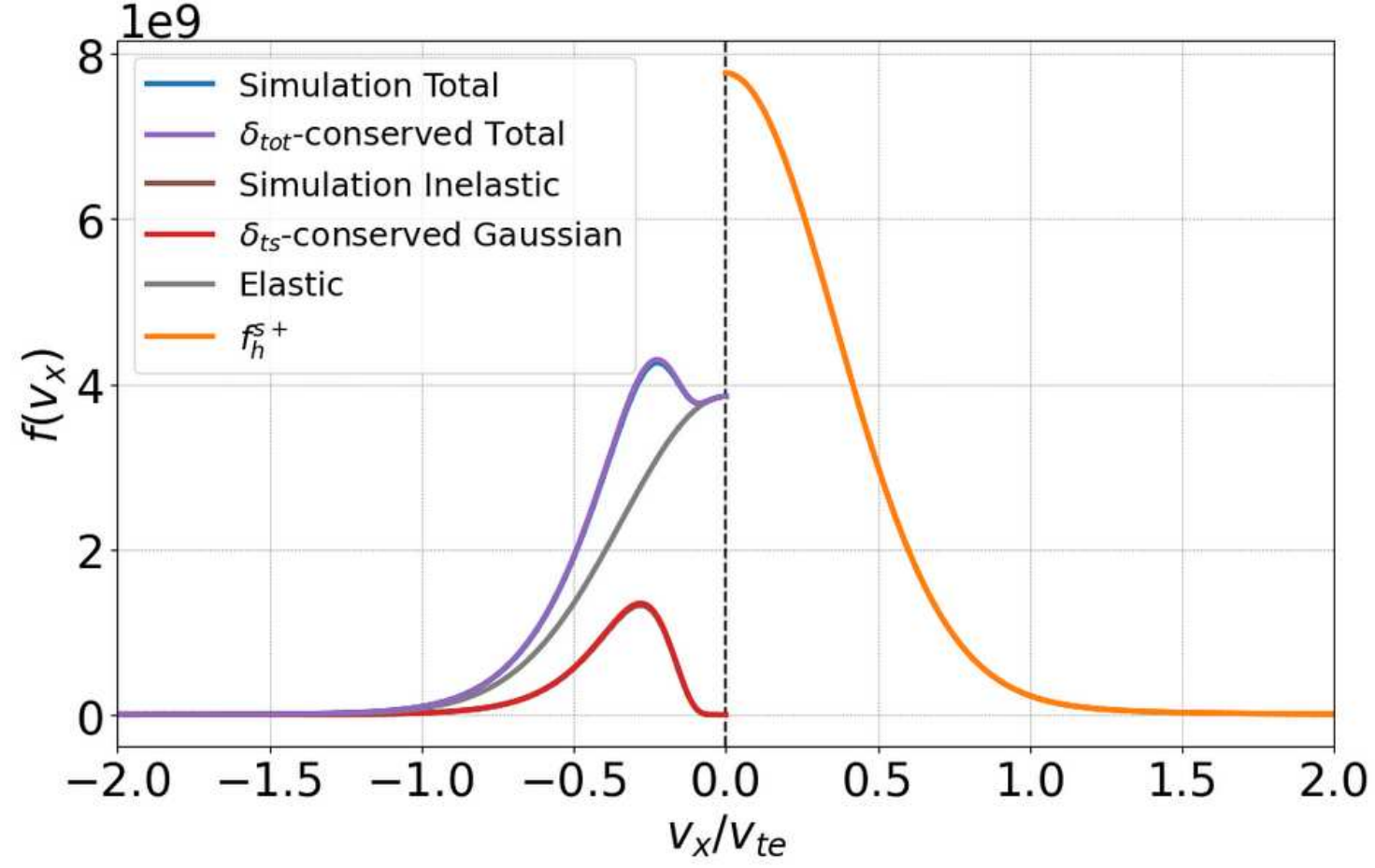}
    \includegraphics[width=0.49\linewidth]{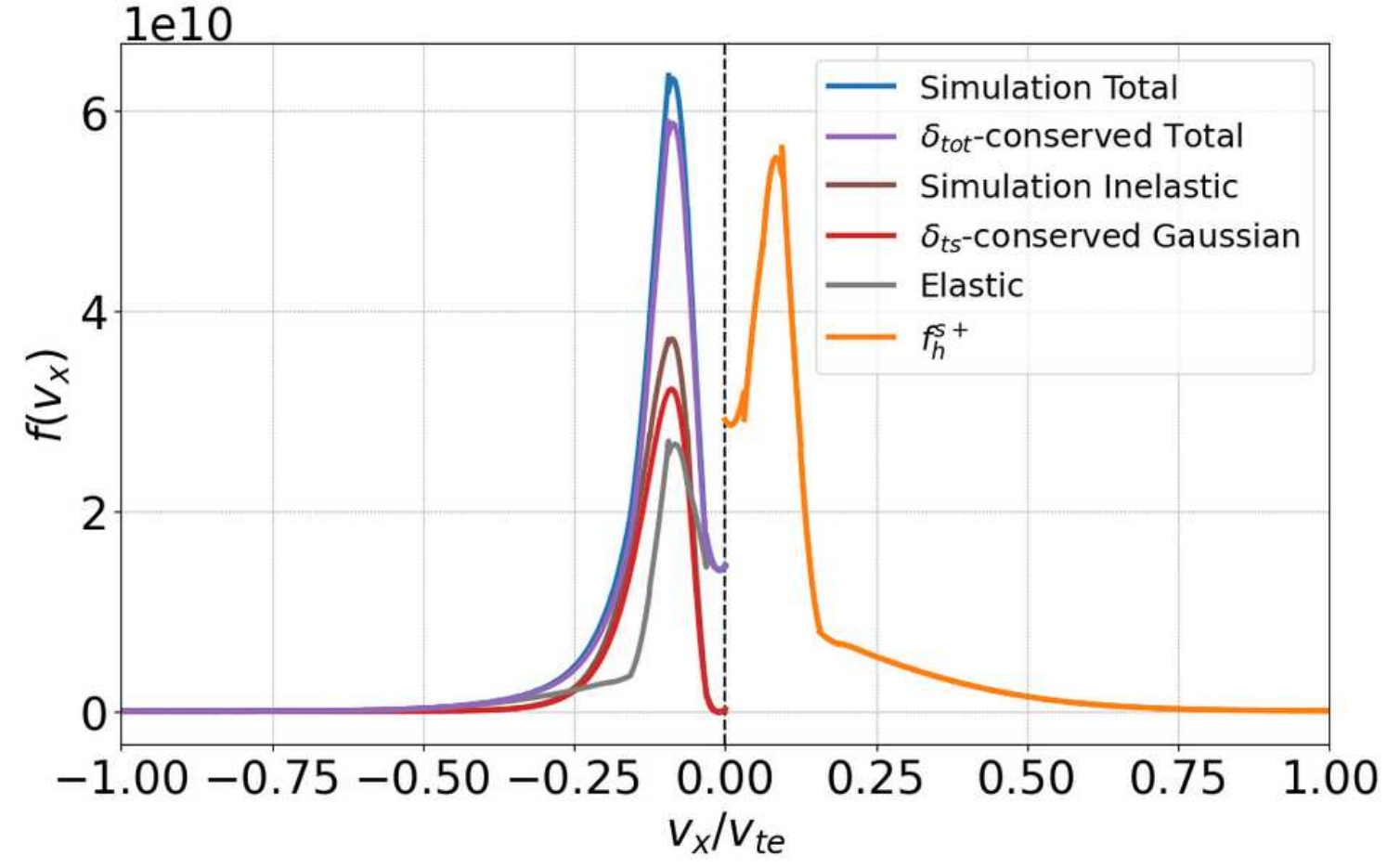}
    \caption{\label{fig:wall_distribution} Impacting and emitted distributions at time $t\omega_{pe}=10000$ for the $T_0=\SI{50}{ev}$ (left, classical) and $T_0=\SI{500}{ev}$ (right, SCL) cases. Also plotted is the full cumulative $\delta_{ts}$-conserving Gaussian calculated over the incoming distribution. In the classical case where the incoming distribution remains Maxwellian, the simulation and conserved curves are closely matched. In the non-Maxwellian SCL distribution, there is a slight drop in accuracy.}
\end{figure*}

\subsection{Simulations}

Test simulations are performed to demonstrate the application of these models and verify that we recover the expected sheath behavior when running in the appropriate yield regime. The Gkeyll code is used for the simulations performed in this work, following the numerical framework described in Section III. For these simulations, the Gaussian spectrum is used with $E_0 = 1.97$ and $\tau = 0.88$. A Maxwellian with $n_0 = \SI{1.0e17}{m^{-3}}$ initial distribution is discretized onto a grid spanning $[0, 128\lambda_D]$ with $N_x = 1024$ in configuration space, and $[-4v_{t,e}, 4v_{t,e}]$ and $[-3u_B, 3u_B]$ for electrons and ions, respectively, with $N_v = 128$ in velocity space, where $v_{t,e} = \sqrt{T_e/m_e}$ is the electron thermal velocity. A mass ratio of $m_i/m_e = 1836$ is used, along with an electron-ion temperature ratio of $T_e/T_i = 1$. Seven cases are performed, with initial temperatures spanning from a $\SI{20}{eV}$-$\SI{2}{keV}$ (see Table~\ref{tb:simulation_results}). LBO collisions are utilized, with a mean free path of $\lambda_{mfp} = 500\lambda_D$, for a few collisions per transit time. Self-species collision frequencies are set to $\nu_{ss} = v_{t,s}/\lambda_{mfp}$, and cross-species to $\nu_{ei} = \nu_{ee}$, $\nu_{ie} = (m_e/m_i)\nu_{ee}$ \cite{braginskii1965}. Finally, a source term is added to the RHS of Eq.~\ref{eq:vlasov} to replenish particle losses to the wall. This is done by scaling the lost ion flux by a linear profile in the presheath and adding equal numbers of electrons and ions back to the simulation,
\begin{equation}
    \frac{\Gamma_i}{L_{src}}\frac{2(L_{src} - x)}{L_{src}}f_{0,s},\quad 0\leq x \leq L_{src},
\end{equation}
where $L_{src} = 40\lambda_D$ is the source length, and $f_{0,s}$ is the Maxwellian at initial temperature normalized to a density of unity.\par
If $\delta_0$ is held constant, we would expect a classical sheath to form for low temperature cases, and ultimately a space-charge limited or inverse sheath to form for high temperature cases. Fig.~\ref{fig:sim_potential} shows that the four lower cases produce a classical sheath, while the three highest temperature cases produce an SCL sheath. Values of the potential at the wall are given in Table~\ref{tb:simulation_results}. The location where the ion drift speed matches the Bohm speed calculated from Eq.~\ref{eq:bohm} is indicated by the dotted vertical lines, with this being further from the wall in the SCL cases than in the classical ones. Very early in time, all cases except the lowest one produce an SCL, however, all but the hottest three transition back to a classical sheath by the time steady-state is achieved. The plot of yield with time in Fig.~\ref{fig:scl_behavior} demonstrates why this is the case, as sheath formation leads to a near-immediate drop in the yield. As the SCL forms, low energy particles emitted from the wall are reflected back to the wall by the potential barrier, leading to a feedback loop of particle accumulation in the distribution at low energy as many of these particles are reemitted in turn due to backscattering. This shifts the balance of the distribution towards zero, driving the total yield lower over time as backscattering begins to dominate over the true secondary emission. This combined with cooling in the sheath eventually drive the yield below the critical value for some cases, leading to the ultimate transition of the SCL sheath to classical profiles.\par
Theory from \cite{taccogna2004} estimates the critical yield to be
\begin{equation}
    \delta_c \approx 1-8.3\sqrt{\frac{m_e}{m_i}}.
\end{equation}
For the simulation's mass ratio of 1/1836, this gives $\delta_c \approx 0.8$. This estimate matches closely the simulation results, as Fig.~\ref{fig:scl_behavior} shows the three cases above this value are all SCL sheaths, while the four below it are classical. Additionally, the two cases that start above and drop below over time are the ones which form an initial SCL which transitions to classical around the same time they fall below this threshold.\par
The simulations reach a steady-state with time. The steady-state presheath temperature, $T_f$ (see Table~\ref{tb:simulation_results}) is significantly colder than the initial temperature. This cooling occurs due to a combination of factors. Sheath formation leads to cooling of the electrons due to the potential field, which reflects some electrons back to the presheath and drives the bulk temperature lower through thermalization. Additionally, the high-energy tail of the distribution is lost to the wall, and while the source replenishes the particle losses, an energy balance is not maintained.
These classical profiles for both cases do appear to trend towards steady-state at long time scales.\par
Fig.~\ref{fig:scl_behavior} also shows the steady-state density profiles for electrons and ions. The classical cases exhibit the standard positive space charge region expected in a sheath. For the SCL cases, the strong emission at the wall leads to a spike in the electron density, which dominates over the ion charge.
None of the cases transition to a reverse sheath, something unsurprising given the lack of a cold ion source in these simulations. While collisions are present, ion collisions are dominated by self-species collisions which do not create substantial amounts of cold ions in the sheath region. Extension of these simulations to capture inverse sheath physics would likely require the addition of either ion-neutral collisions or a source term in the sheath. Detailed analysis of the inverse sheath regime, emission feedback mechanisms, and how the space-charge limited and inverse sheath regimes are impacted by the dynamic emission, particularly in comparison to the constant case, will be examined in future work. For now the main focus of these cases is to demonstrate that the features of emission behavior are highly dependent on energy-dependent mechanisms, and that we do indeed recover predicted sheath behavior in corresponding yield regimes.\par
The electron distributions are plotted in full phase space in Fig.~\ref{fig:sim_dist} for a representative classical case ($T_0=\SI{50}{eV}$) and SCL case ($T_0=\SI{500}{eV}$). The major feature evident in the SCL case is an emitted electron beam which is accelerated off the wall, before the collisions diffuse it as it propagates into the presheath. Fig.~\ref{fig:wall_distribution} shows the steady-state electron distribution function at the wall for the same representative cases. In the classical case, electrons continue to impact the wall in a Maxwellian distribution. For the SCL, however, the potential barrier accelerates the population, leading to a double-peaked distribution very close to the surface. The emitted distribution is decomposed into the component populations of inelastically emitted and elastically backscattered particles, and compared with the complete summation similar to that in Fig.~\ref{fig:see_spectra} of $\delta_{ts}$-conserving Gaussian distributions across the incoming distribution energies. The results show that as long as the incoming distribution remains Maxwellian, the calculation of $\bar{\delta}$ remains highly accurate and the theory and simulation predictions of the spectrum match closely. However, the non-Maxwellian distributions found in the SCL cases cause the weighted averaging to lose some accuracy, with Table~\ref{tb:simulation_results} showing errors of 5-10\% introduced for the estimation of $\delta_{tot}$ in the cases studied. The highest temperature case sees a spike in the error to around 20\% due to extremely poor resolution of the distribution function on the larger domain. Increased velocity space resolution causes this error to decrease, as it seems to primarily be caused by sharp gradients within a cell yielding a poor average. This accuracy issue does not extend to the elastic backscattering.

\begin{table}
\centering
\begin{tabular}{ |c|c|c|c|c|  }
\hline
$T_0$ & $T_f$ & $\delta_{tot}$ & $\tilde{\phi}_w$ & $\Delta\delta_{tot}$ \\
\hline
$\SI{20}{eV}$ & $\SI{2.54}{eV}$ & $0.53$ & $-2.78$ & 1.9\%\\
$\SI{50}{eV}$ & $\SI{6.44}{eV}$ & $0.59$ & $-2.04$ & 0.7\%\\
$\SI{100}{eV}$ & $\SI{11.57}{eV}$ & $0.67$ & $-1.82$ & 1.0\%\\
$\SI{200}{eV}$ & $\SI{19.53}{eV}$ & $0.75$ & $-1.45$ & 1.1\%\\
$\SI{500}{eV}$ & $\SI{41.78}{eV}$ & $0.84$ & $-1.26$ & 8.0\%\\
$\SI{1000}{eV}$ & $\SI{80.68}{eV}$ & $0.89$ & $-1.36$ & 6.9\%\\
$\SI{2000}{eV}$ & $\SI{160.22}{eV}$ & $0.83$ & $-1.47$ & 20.1\%\\
\hline
\end{tabular}
\caption{Simulation case results. Initial and steady-state temperature, electron yield, normalized wall potential ($\tilde{\phi}_w=q_0\phi_w/T_f$), and yield error are given.}
\label{tb:simulation_results}
\end{table}

\section{Conclusions}

Presented in this work is an energy-dependent modeling of secondary electron emission in a continuum kinetic code. Both the elastic and inelastic populations are implemented based on data-driven models. The classical or SCL sheath modes match well to theoretical predictions for the critical yield, with high-$\delta$ cases transitioning to a space-charge limited sheath.\par
While the emission model presented here is demonstrably robust, additional consideration must be taken in future work to produce rigorously accurate and predictive simulations. No consistent mechanism for cold ions in the sheath is currently present which would allow the sheath to transition to the inverse regime. The plasma regime changes over the course of the simulation due to cooling mechanisms, as the source term maintains only the particle balance, and does not maintain an energy balance. While a steady-state can still be achieved, it is at a drastically lower temperature than the initial state, something undesirable if attempting to make sheath predictions applicable to particular devices by matching the parameter regime. Finally, the simulations presented here are limited to a single velocity space dimension, not accounting for the angular dependence of the emission. The theoretical extension of the emission model is supplied for 2V, however the resolution required to efficiently simulate multiple velocity dimensions is still currently prohibitive. The resolution currently used in the simulations is necessary to capture the near-wall region of configuration space and the low-energy region of velocity space where most of the sharp features of the emission are concentrated. As the grid is uniform, this means that the rest of the domain tends to have higher resolution than is necessary, dragging down the computational performance. Implementation of non-uniform grids will likely be necessary moving forward to maintain resolution in important regions while avoiding the current over-resolution of the presheath and high-velocity parts of the domain.\par
This work presents the first description of a self-consistent, physically-representative model of electron emission from material walls for a range of plasma regimes in continuum kinetics.

\begin{acknowledgments}
The work presented here was supported by the U.S. Department of Energy ARPA-E BETHE program under Grant No. DE-AR0001263. The authors acknowledge Advanced Research Computing at Virginia Tech for supplying computational resources and support for this work.
\end{acknowledgments}

\section*{Data Availability Statement}

All the simulation results presented in this paper were produced by and are reproducible using the open-source \texttt{Gkeyll} software. Information for obtaining, installing, and running \texttt{Gkeyll} may be found on the documentation site \cite{gkyldocs}. The input files for the simulations used to produce the results in this paper may be acquired from the repository at \url{https://github.com/ammarhakim/gkyl-paper-inp/tree/master/2023_PSST_ModelSEE}.
\nocite{*}

\bibliography{apssamp}

\end{document}